\documentclass[graybox]{svmult}

\usepackage{mathptmx}       % selects Times Roman as basic font
\usepackage{helvet}         % selects Helvetica as sans-serif font
\usepackage{courier}        % selects Courier as typewriter font
\usepackage{type1cm}        % activate if the above 3 fonts are
\usepackage{makeidx}         % allows index generation
\usepackage{graphicx}        % standard LaTeX graphics tool
\usepackage{multicol}        % used for the two-column index
\usepackage[bottom]{footmisc}% places footnotes at page bottom

\usepackage[numbers,sort&compress]{natbib}

\makeindex             % used for the subject index
\begin{document}

\title*{Generation of defects and disorder from deeply quenching a liquid to form a solid}
%\titlerunning{Generation of defects and disorder from deeply quenching a liquid to form a solid}
% Use \titlerunning{Short Title} for an abbreviated version of
% your contribution title if the original one is too long
\author{A.~J.~Archer, M.~C.~Walters, U.~Thiele and E.~Knobloch}
% Use \authorrunning{Short Title} for an abbreviated version of
% your contribution title if the original one is too long
\institute{A.~J.~Archer \at Department of Mathematical Sciences, Loughborough University, Loughborough LE11 3TU, UK \email{a.j.archer@lboro.ac.uk}
\and M.~C.~Walters \at Department of Mathematical Sciences, Loughborough University, Loughborough LE11 3TU, UK \email{m.walters@lboro.ac.uk}
\and U.~Thiele \at Institut f\"ur Theoretische Physik, Westf\"alische Wilhelms-Universit\"at M\"unster, Wilhelm Klemm Str.\ 9, D-48149 M\"unster, Germany \\ Center of Nonlinear Science (CeNoS), Westf\"alische Wilhelms Universit\"at M\"unster, Corrensstr. 2, 48149 M\"unster, Germany\\ \email{u.thiele@uni-muenster.de}
\and E.~Knobloch \at Department of Physics, University of California at Berkeley, Berkeley, California 94720, USA \email{knobloch@berkeley.edu}}
%
% Use the package "url.sty" to avoid
% problems with special characters
% used in your e-mail or web address
%
\maketitle

\abstract{We show how deeply quenching a liquid to temperatures where it is linearly unstable and the crystal is the equilibrium phase often produces crystalline structures with defects and disorder. As the solid phase advances into the liquid phase, the modulations in the density distribution created behind the advancing solidification front do not necessarily have a wavelength that is the same as the equilibrium crystal lattice spacing. This is because in a deep enough quench the front propagation is governed by linear processes, but the crystal lattice spacing is determined by nonlinear terms. The wavelength mismatch can result in significant disorder behind the front that may or may not persist in the latter stage dynamics. We support these observations by presenting results from dynamical density functional theory calculations for simple one- and two-component two-dimensional systems of soft core particles.}

\section{Introduction}
\label{sec:1}

Solids with a well-ordered crystalline structure have numerous applications in materials science. In this article we focus on some of the considerations that determine whether a crystal grown from a supercooled liquid will be essentially defect-free or will contain substantial disorder that will modify its material properties. We are interested in particular in the consequence of quenching a uniform liquid to a temperature at which (a) the thermodynamically stable state is a crystalline solid, and (b) the supercooled liquid is linearly unstable with respect to growing density perturbations. This occurs when the temperature quench is sufficiently deep. When the temperature of the liquid is quenched only a little below the freezing temperature, the crystal forms instead via nucleation and growth \cite{debenedetti_book, oxtoby1992, oxtoby1998, sear2007}. However, if the temperature quench is sufficiently deep, then the uniform liquid is unstable and any small perturbations in the density grow spontaneously \cite{TDP06, KL86, AWTK14}. This is often referred to as the spinodal regime. One expects that in this case the perturbations will evolve into a well-ordered crystalline solid but whether this is the case depends crucially on the speed of the crystallisation front that develops from an initial small amplitude density perturbation \cite{AWTK14, ARTK12}. Of course we anticipate that the front speed will be slow for shallow quenches and faster for deep quenches, an expectation we confirm here. The front speed plays a crucial role since the wavenumber of the density modulations deposited behind the crystallisation front is in fact determined dynamically and so may differ from the wavenumber of the crystal in thermodynamic equilibrium. We examine here the physical mechanisms responsible for the speed of the crystallisation front and distinguish between the so-called pulled fronts whose speed is determined by linear processes and pushed fronts whose speed is determined nonlinearly \cite{saarloos}. 

The fact that the wavelength of the density modulations deposited by a pulled crystallisation front differs in general from the length scale of the equilibrium crystal is important in determining the nature of the solid structure that is formed. If the length scales are similar, then behind the front few defects are generated as the solid relaxes to equilibrium. In contrast, if the resulting wavenumber mismatch is large, then the defect density and disorder in the solid are greatly enhanced. We find in general that this is particularly so for deep quenches where the front speed is high.

We elucidate these notions by studying the solidification of model two-dimensional (2D) systems of soft-core particles. We use a simple mean-field dynamical density function theory (DDFT) \cite{MaTa99, MaTa00, ArEv04, ArRa04, HM} that has been shown to be quite accurate in reproducing Brownian dynamics computer simulations to demonstrate that a mismatch between the wavelength of the modulations deposited behind a solidification front and the length scale of the equilibrium crystal is indeed likely when the liquid is deeply quenched. We also observe that this mismatch results in a disordered structure behind the front. In systems composed of only one species of particles, the disorder subsequently largely disappears as the particles are able to rearrange, healing the majority of the defects. However, in binary mixtures this disorder may be frozen-in. 

{In this paper we briefly review the pertinent results from our earlier work \cite{AWTK14}, and then focus on characterizing the disorder created by fast crystallisation fronts in several systems of interest and the factors influencing whether the resulting disorder persists over long time scales. We emphasise that wavenumber selection} via a moving front is a linear process only when the temperature quench is deep. Shallow quenches to a temperature just below where the liquid and crystalline phases coexist do not generate crystallisation fronts unless the crystal phase is nucleated; the resulting front is necessarily a pushed front. Furthermore, this type of fronts persists into the linearly unstable state until the quench becomes so deep that the speed of the pulled front exceeds the speed of the pushed front \cite{AWTK14}.

Propagation of soft matter crystallisation fronts has been considered before, notably in Refs. \cite{LG89,GaEl11,ARTK12} within the so-called phase-field crystal description in one spatial dimension and in Ref.\ \cite{CGOP94, TBVL09, ELWRGTG12} in two spatial dimensions, in both cases focusing on the properties of pulled fronts. The more accurate DDFT approach used here extends and generalises these results to both types of crystallisation fronts and to two spatial dimensions. 

We should mention that the DDFT that we use assumes that the particles in the system follow {overdamped but} stochastic equations of motion -- i.e.\ they are implicitly treated as colloidal particles immersed in a solvent that acts as a heat bath. Thus, after the quench the system is maintained at the quench temperature. This contact with a heat bath eliminates the effects of latent heat release. {The absence of particle inertia implies that the dynamics are diffusive: no propagating excitations (phonons) are possible in such systems.}

This paper is structured as follows: In \S\ref{sec:2} and \ref{sec:3} we describe the origin of the basic length scales involved in these processes. In \S\ref{sec:4} and \ref{sec:5} we summarise and illustrate the basic theory determining the speed of pulled fronts. In \S\ref{sec:6} and \S\ref{sec:7} we describe the DDFT theory for a single component fluid and the structures generated by crystallisation fronts moving with different speeds. \S\ref{sec:8} and \S\ref{sec:9} compare and contrast these results with the results obtained for binary mixtures. The article concludes with a brief summary and a few concluding remarks.

\section{Length scales in liquids and solids}\label{sec:2}

Consider a system of atoms, molecules, colloids, etc (henceforth referred to as `particles') which collectively exhibit a phase transition from a liquid to a crystalline state. Thermodynamic and structural properties of these two phases can in principle be found using classical density functional theory (DFT) \cite{Evans79, Evans92, lutsko10, HM}. In DFT it is shown that there exists a functional $\Omega[\rho]$, together with a minimisation principle
\begin{equation}\label{eq:min_principle}
\frac{\delta \Omega[\rho]}{\delta \rho}=0.
\end{equation}
The density profile $\rho^*({\bf r})$ that solves this equation, i.e.\ that minimises $\Omega[\rho]$, is the density distribution of the system at equilibrium. Furthermore, $\Omega[\rho^*]$ is the thermodynamic grand potential of the system. Solving Eq.~(\ref{eq:min_principle}) for state points in the phase diagram where the liquid is the equilibrium phase yields a density profile that is uniform in space, $\rho({\bf r})=\rho_0$. In contrast, for the crystal phase, the density profile exhibits a regular array of peaks. From this density profile, quantities such as the crystal lattice spacing $\ell$ can be determined.

The functional $\Omega[\rho]$ is highly nonlinear, and quantities characterising the crystal, such as $\ell$, depend on all terms in the functional. In contrast, quantities such as the static structure factor $S(k)$ \cite{HM} of the liquid only depend on the linear response of the liquid and so only depend on the terms in $\Omega[\rho]$ that are quadratic in the density fluctuation $\tilde{\rho}\equiv\rho-\rho_0$. Related to the static structure factor is the Fourier transform of the linear density response function $\chi({\bf r}-{\bf r}')$, viz., $\hat{\chi}(k)=-(\rho_0/k_BT)S(k)$, that relates the change in the density $\delta \rho({\bf r})$ to a change $\delta V_{ext}({\bf r})$ in the external potential \cite{Evans79}:
\begin{equation}\label{eq:linear_response}
\delta \rho({\bf r})=-\int d{\bf r}'\chi({\bf r},{\bf r}')\delta V_{ext}({\bf r}').
\end{equation}
This formula applies for both uniform and non-uniform fluids; in particular, in the case of a uniform fluid with density $\rho_0$ and $V_{ext}=0$ perturbed by a small amplitude external potential $\delta V_{ext} ({\bf r})$, Eq.\ (\ref{eq:linear_response}) determines the resulting change in the density profile: $\delta\rho\equiv\rho-\rho_0=\tilde{\rho}$.\footnote{The result in Eq.\ (\ref{eq:linear_response}) also applies to non-uniform liquids, i.e.\ to liquids initially at equilibrium with a density profile $\rho_{old}({\bf r})$ in an external potential $V_{old}({\bf r})$, disturbed by an infinitesimal change to the external potential, $V_{old}({\bf r})\to V_{new}({\bf r})$. The resulting change in the density profile $\delta\rho({\bf r})\equiv\rho_{new}({\bf r})-\rho_{old}({\bf r})$ is then also given by Eq.\ (\ref{eq:linear_response}), where $\delta V_{ext}({\bf r})\equiv V_{new}({\bf r})-V_{old}({\bf r})$.}

The main point of the above comments is to emphasise that quantities pertaining to the crystal, such as $\ell$, depend on all terms in $\Omega[\rho]$, but quantities such as $S(k)$ and the dispersion relation $\omega(k)$ \cite{ArEv04, ARTK12, AWTK14}, only depend on the quadratic terms in $\tilde{\rho}$. We emphasise this point because when a uniform liquid is deeply quenched, the length scales of the density modulations that initially grow after the quench are determined by $\omega(k)$ and so only depend on the quadratic terms in $\tilde{\rho}$. In particular, the principal peak in the dispersion relation $\omega(k)$ determines the wavenumber of the density fluctuation that grows the fastest. The fact that this wavenumber is determined by a quantity that only depends on the quadratic terms in $\tilde{\rho}$ shows that these fastest growing modes need not have the equilibrium wavenumber $2\pi/\ell$, i.e.\ they do not necessarily generate the correct density modulations for a perfect equilibrium crystal. From these considerations, one can infer that a deeply quenched liquid may well produce a disordered solid, because the length scale of the fastest growing modes is not in general equal to $\ell$. This argument does not address whether, as solidification proceeds, the system can rearrange and subsequently anneal all the defects generated in the initial stages of the solidification process to produce a perfect crystal. Nonetheless, the observation that the initial dynamics after the quench do not in general produce density fluctuations of the correct length scale is an important observation.

\section{Dispersion relation}\label{sec:3}

The time evolution of the density distribution $\rho({\bf r},t)$ in a fluid system of particles is given by the continuity equation
\begin{equation}\label{eq:continuity}
\frac{\partial\rho}{\partial t}=-\nabla\cdot{\bf j},
\end{equation}
where ${\bf j}$ is the current. This equation is simply a statement of the fact that the number of particles in the system is a conserved quantity. To solve this equation of course requires an expression for the current ${\bf j}\equiv\rho {\bf u}$, where ${\bf u}$ is the local fluid velocity. In general, we only have formal expressions for this quantity, and to actually calculate the fluid dynamics, approximations are required. For example, for colloidal fluids, where the particles move with stochastic (Brownian) equations of motion, the following approximation can be rather good:
\begin{equation}\label{eq:DDFT_current}
{\bf j}=-\Gamma\rho\nabla\frac{\delta \Omega[\rho]}{\delta \rho},
\end{equation}
where $\Gamma=D/k_BT$ is a mobility coefficient, $D$ is the diffusion coefficient, $k_B$ is Boltzmann's constant and $T$ is the temperature. This approximation is the central result in DDFT \cite{MaTa99, MaTa00, ArEv04, ArRa04, HM}.

Consider now a fluid with uniform density $\rho_0$ with a superposed small amplitude density perturbation $\tilde{\rho}\equiv\rho-\rho_0$ that is either artificially imposed, for example, via an external field or a consequence of random thermal fluctuations.  Eq.\ (\ref{eq:continuity}) shows that the perturbation evolves according to
\begin{equation}\label{eq:continuity_lin}
\frac{\partial\tilde{\rho}}{\partial t}={\cal L}\tilde{\rho}+{\cal O}(\tilde{\rho}^2),
\end{equation}
where ${\cal L}$ is an operator that is obtained by linearising the full dynamical equation (\ref{eq:continuity}). For example, in the colloidal case, where Eq.\ (\ref{eq:DDFT_current}) is applicable, the operator ${\cal L}=D \nabla^2-D\rho_0 \nabla^2c^{(2)} \otimes$, where $\otimes$ denotes a convolution, i.e.\ $c^{(2)} \otimes\tilde{\rho}\equiv\int d{\bf r}' c^{(2)}({\bf r}-{\bf r}')\tilde{\rho}({\bf r}')$, and $c^{(2)}({\bf r})$ is the Ornstein-Zernike pair direct correlation function \cite{ArEv04, HM}. Linearising Eq.\ (\ref{eq:continuity_lin}) and decomposing $\tilde{\rho}$ into a sum of different Fourier modes,
\begin{equation}\label{eq:Fourer_modes}
\tilde{\rho}({\bf r},t)=\sum_{\bf k}\hat{\rho}_{\bf k}e^{i{\bf k}\cdot{\bf r}+\omega(k)t},\qquad k\equiv|{\bf k}|,
\end{equation}
{leads to a dispersion relation for the growth rate $\omega(k)$ of density fluctuations with wavenumber $k$.} In the colloidal case described by Eq.\ (\ref{eq:continuity_lin}) we obtain \cite{ArEv04, ARTK12, AWTK14}:
\begin{equation}\label{eq:lin}
\omega(k)=-Dk^2[1-\rho_0\hat{c}(k)],
\end{equation}
where $\hat{c}(k)$ is the Fourier transform of $c^{(2)}(r)$. Note that for the equilibrium fluid, we also have the relation $S(k)=[1-\rho_0\hat{c}(k)]^{-1}$.

\begin{figure}%[t]
\sidecaption[t]
\includegraphics[scale=1.]{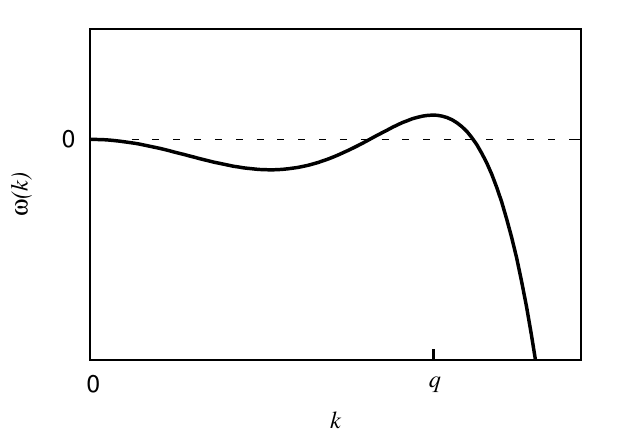}
\caption{Typical form of the dispersion relation $\omega(k)$ for a uniform liquid that is supercooled to the region of the phase diagram where it is linearly unstable against the growth of density modulations with wavenumber $k\approx q$.}
\label{fig:disp_rel_sketch}
\end{figure}

Equation (\ref{eq:Fourer_modes}) tells us that wavenumbers $k$ for which $\omega(k)>0$ grow while those for which $\omega(k)<0$ decay. For a deeply supercooled liquid, it can be the case that $\omega(k)$ is positive for a band of wavenumbers around the value $k=q$, say. A sketch of the typical form of $\omega(k)$ for such a supercooled liquid is displayed in Fig.\ \ref{fig:disp_rel_sketch}.

Thus in the early stages following a deep quench, we see the growth of density modulations with wavenumber $k\approx q$, leading to the appearance of the length scale $2\pi/q$ in the density distribution of the system. However, as emphasised above, this scale may differ from the equilibrium crystal lattice spacing $\ell$.

\section{Speed of solidification fronts: {marginal stability hypothesis}}
\label{sec:4}

A further aspect that has not been mentioned in the discussion so far relates to the question of how do solidification fronts propagate into the unstable liquid and what length scale density modulations do such solidification fronts produce? In the case where the liquid is unstable the properties of the solidification front can often be determined from the marginal stability hypothesis \cite{DL83, BBDKL85, GaEl11, ARTK12, AWTK14}: Consider the leading edge of such a front, where the growing density modulations are still small in amplitude and suppose this front is advancing with velocity ${\bf v}$. In a reference frame that moves with the front, Eq.\ (\ref{eq:continuity_lin}) becomes
\begin{equation}\label{eq:continuity_lin_moving}
\frac{\partial\tilde{\rho}}{\partial t}+{\bf v}\cdot \nabla\tilde{\rho}={\cal L}\tilde{\rho}+{\cal O}(\tilde{\rho}^2)
\end{equation}
{and so the (complex) growth rate of density perturbations in this moving reference frame becomes $\omega_{\bf v}(k)=i{\bf k}\cdot{\bf v}+\omega(k)$. The marginal stability hypothesis posits that the density modulation at the leading edge of the front has zero growth rate in the comoving frame -- if this mode were to have a positive growth rate, the front would be moving faster than the reference frame; if it were to have a negative growth rate the front would retract. In other words, the (complex) group velocity in the comoving reference frame must be zero and the growth rate of the density perturbations must also be zero, i.e.,}
\begin{equation}\label{eq:front1}
iv+\frac{d\omega(k)}{dk}=0
\end{equation}
and
\begin{equation}\label{eq:front2}
Re[i{\bf k}\cdot{\bf v}+\omega(k)]=0,
\end{equation}
where $k\equiv k_r+ik_i$ and $v=|{\bf v}|$. {These equations can in fact be derived, under appropriate conditions, using the method of stationary phase applied to the longtime evolution of an infinitesimal spatially localized initial density perturbation \cite{saarloos}.} Together they provide three conditions that are to be solved for the three unknowns, $v$, $k_r$ and $k_i$. In the front region, the density profile $\tilde{\rho}({\bf r},t)$ takes the form $\rho_{\rm front}({\bf \xi},t)$, where ${\bf \xi}\equiv {\bf r}-{\bf v}t$, since the moving front generates wavelengths in a periodic fashion.  Without loss of generality we take the front to move along the $x$-axis, i.e.\ ${\bf v}=(v,0,0)$, implying that $\tilde{\rho}({\bf r},t)\sim \exp(-k_ix)\sin(k_r(x-vt)+Im[\omega(k)]t)$. Thus $k_r$ determines the wavelength of the density modulations in the front. More importantly, if no phase slips take place, then the wavenumber of the density modulations left behind by the front is \cite{BBDKL85, GaEl11, ARTK12, AWTK14}:
\begin{equation}\label{eq:k_star}
k^*=k_r+\frac{1}{v}Im[\omega(k)].
\end{equation}
Thus $k^*$ is determined by the form of $\omega(k)$, which is obtained from the linearised equation (\ref{eq:continuity_lin}). Therefore the wavelength $2\pi/k^*$ of the density modulation created behind the advancing front also differs in general from the equilibrium crystal lattice spacing $\ell$, just like the wavelength associated with the fastest growing mode.

Thus even if the length scale determined by the maximum in $\omega(k)$ is the same as the equilibrium lattice spacing for the crystal, as is the case in the simple PFC theory for the crystal, solidification fronts advancing into a deeply supercooled liquid will still generate density modulations with a distinct wavelength, requiring substantial subsequent rearrangements of the system in order to form a defect-free crystal without strain.

\section{Marginal stability calculation for simple model}\label{sec:5}

In this section we perform the marginal stability calculation to obtain the front speed $v$ and wavenumber $k^*$ of the density modulations created behind the front to show how these quantities depend on the dispersion relation. We approximate the dispersion relation by making a Taylor expansion around the wavenumber corresponding to the principal peak and truncating after the $k^3$ term,
\begin{equation}\label{eq:disp_rel_approx}
\omega(k)\approx\omega_m-a(k-q)^2-b(k-q)^3,
\end{equation}
where $\omega_m=\omega(k=q)$ is the maximum growth rate. The coefficient $a>0$ is related to the width of the principal peak while $b$ measures its asymmetry around the peak wavenumber $k=q$. 

Substituting Eq.\ (\ref{eq:disp_rel_approx}) into Eq.\ (\ref{eq:front1}) we obtain:
\begin{equation}
iv-2a(d+ik_i)-3b(d+ik_i)^2=0,
\end{equation}
where we have written $d=k_r-q$. Separating the real and imaginary parts of this equation, we obtain the following expressions for the front speed and the imaginary part of $k$:
\begin{eqnarray}
v=2(a+3bd)k_i\label{eq:v}\\
k_i=\sqrt{\frac{2ad}{3b}+d^2}\label{eq:k_i}.
\end{eqnarray}
Substituting Eq.\ (\ref{eq:disp_rel_approx}) into Eq.\ (\ref{eq:front2}) we obtain:
\begin{equation}
Re[iv(k_r+ik_i)+\omega_m-a(d+ik_i)^2-b(d+ik_i)^3]=0
\end{equation}
giving
\begin{equation}\label{eq:other}
-vk_i+\omega_m-a(d^2-k_i^2)-b(d^3-3dk_i^2)=0.
\end{equation}
Inserting Eqs.\ (\ref{eq:v}) and ({\ref{eq:k_i}) into Eq.\ (\ref{eq:other}) we obtain a cubic equation to be solved for $d$. However, for present illustrative purposes it is instructive to proceed analytically on the assumption that $d$ is a small quantity. In this case Eqs.\ (\ref{eq:v}) and ({\ref{eq:k_i}) become
\begin{eqnarray}
v\approx2ak_i\label{eq:v_smalld}\\
k_i\approx\sqrt{\frac{2ad}{3b}}\label{eq:k_i_smalld}.
\end{eqnarray}
Linearization of Eq.\ (\ref{eq:other}) in $d$ now leads to $d=3b\omega_m/2a^2$, i.e.\ to
\begin{equation}\label{eq:k_r_approx}
k_r\approx q+\frac{3b\omega_m}{2a^2}.
\end{equation}
This result shows that the wavenumber $k_r$ of the density modulation in the advancing solidification front is not equal to the wavenumber of the fastest growing mode for the quenched uniform fluid, $q$, unless the peak of the dispersion relation is symmetric, i.e.\ unless $b=0$. We also see that the difference between these two wavenumbers grows with $\omega_m$, the magnitude of which is related to the degree of undercooling. The deeper the quench, the larger is $\omega_m$. Moreover, inserting these results into Eq.\ (\ref{eq:k_star}) we obtain the wavenumber $k^*$ of the modulations deposited behind the front:
\begin{equation}\label{eq:k_star_smalld}
k^*\approx q+\frac{b\omega_m}{2a^2}.
\end{equation}
Thus the wavenumber $k^*$ differs in general from the fastest growing wavenumber $q$, and neither of these wavenumbers is in general equal to $2\pi/\ell$ and so defects and disorder must be present shortly after a deep quench. Some systems are subsequently able to rearrange, but others are not, as we show in the subsequent sections for a particular model fluid composed of soft-core particles.

\section{Model fluid}\label{sec:6}

We consider a 2D system of soft-core particles interacting via the so-called generalised exponential model of index $n$ (GEM-$n$) pair-potential:
\begin{equation}\label{eq:pair_pot}
w(r)=\epsilon e^{-(r/R)^n},
\end{equation}
where the parameter $0<\epsilon<\infty$ determines the energy penalty for a pair of particles to overlap completely, $R$ is the radius of the particles and the exponent $n$ determines the `softness' of the potential. When $n=2$, the potential varies slowly. In contrast, when $n$ is large, as the separation distance $r$ between a pair of particles is decreased, the potential increases rapidly from $\approx 0$ to a value $\approx \epsilon$ over a short distance at $r=R$. Here, we consider the case when $n=8$. Such soft potentials arise as the effective interaction potential between polymers or soft macromolecules in solution \cite{Likos01, DaHa94, likos:prl:98, LBHM00, BLHM01, JDLFL01, Dzubiella_2001, LBFKMH02, likos:harreis:02, GHL04, MFKN05, Likos06, LBLM12}.

Our main reason for considering this model centres on the fact that the structure and phase behaviour (i.e.\ thermodynamics) of this model is well described by a rather simple approximation for the free energy. The grand potential of the system can be decomposed as follows \cite{Evans79, Evans92, lutsko10, HM}:
\begin{equation}\label{eq:grand_pot}
\Omega[\rho({\bf r})]=\mathcal{F}[\rho({\bf r})]+\int d{\bf r}\rho({\bf r})(V_{ext}({\bf r})-\mu),
\end{equation}
where $\mu$ is the chemical potential and $\mathcal{F}[\rho({\bf r})]=\mathcal{F}_{id}[\rho({\bf r})]+\mathcal{F}_{ex}[\rho({\bf r})]$ is the intrinsic Helmholtz free energy functional. The ideal gas contribution is
\begin{equation}\label{eq:F_id}
 \mathcal{F}_{id}[\rho({\bf r})]=k_BT \int d{\bf r} \rho({\bf r})\left(\ln[\rho({\bf r})\Lambda^2]-1\right),
\end{equation}
where $\Lambda$ is the thermal de Broglie wavelength and we use the following mean-field approximation for the excess contribution to the free energy \cite{Likos01}: 
\begin{equation}
 \mathcal{F}_{ex}[\rho({\bf r})]=\frac{1}{2}\int d{\bf r} \int d {\bf r}'\rho({\bf r})w(|{\bf r}-{\bf r}'|)\rho({\bf r}')
 \label{eq:F_ex}
 \end{equation}
 which has been widely used in studies of the structure and phase behaviour of soft-core systems \cite{Likos01, ArEv01, ArEv02, AER02,  ALE02, ALE04, GAL06, MGKNL06, MGKNL07, MoLi07, LMGK07, MCLFK08, LMMGK08, OvLi09b, OvLi09, TMAL09, CML10, MaLi11, NKL12, CPPR12, Pini14, ARK13, AWTK14}.
 
 \begin{figure}%[t]
\sidecaption[t]
\includegraphics[scale=0.9]{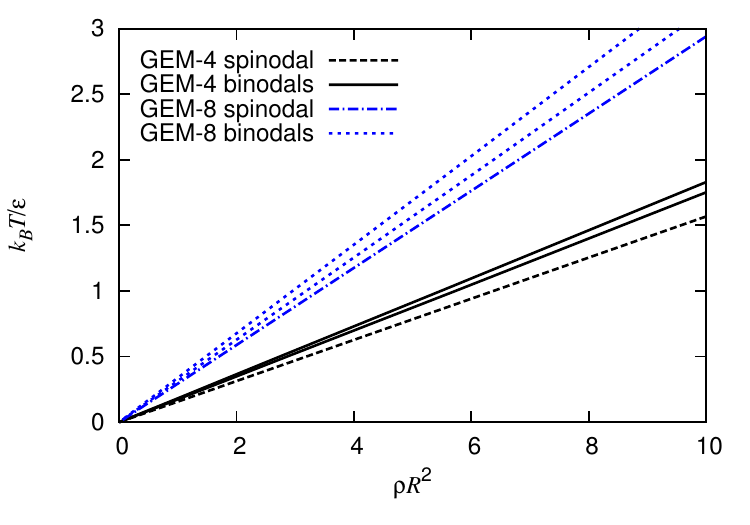}
\caption{The phase diagram of 2D GEM-4 and GEM-8 fluids. The binodals are lines of thermodynamic coexistence; the spinodals correspond to the onset of linear instability of the metastable uniform liquid.}
\label{fig:phase_diag}
\end{figure}

The bulk phase diagrams for the 2D {GEM-4 and GEM-8 systems were calculated in Ref.\ \cite{AWTK14} (see also Ref.\ \cite{PrSa14}) and are displayed in Fig~\ref{fig:phase_diag}.} The diagrams exhibit a liquid phase at low densities and/or high temperatures which freezes to form a novel cluster crystal phase as the temperature decreases or the density increases.

We assume that the particles move with overdamped stochastic (Brownian) equations of motion. Therefore, the time evolution of the non-equilibrium density profile $\rho({\bf r},t)$ may be determined using DDFT, i.e., using Eqs.\ (\ref{eq:grand_pot}) -- (\ref{eq:F_ex}) together with Eqs.\ (\ref{eq:continuity}) and (\ref{eq:DDFT_current}). Within the present mean-field approximation, the pair direct correlation function is \cite{Evans79,Evans92,lutsko10,HM}:
\begin{equation}
c^{(2)}(|{\bf r}-{\bf r}'|)\equiv-\beta\frac{\delta^2 \mathcal{F}_{ex}[\rho({\bf r})]}{\delta \rho({\bf r}) \delta \rho({\bf r}')}=-\beta w(|{\bf r}-{\bf r}'|),
\end{equation}
where $\beta=(k_BT)^{-1}$ and so the dispersion relation (\ref{eq:lin}) has the following very simple form:
\begin{equation}\label{eq:disp_rel}
\omega(k)=-Dk^2[1+\rho_0\beta \hat{w}(k)],
\end{equation}
where $\hat{w}(k)$ is the 2D Fourier transform of the pair potential $w(r)$. The threshold for linear instability of the uniform fluid is determined by $\omega(k=q)=0$, where $q\neq0$ is the wavevector at which $\omega(k)$ has a maximum, i.e., by $1+\rho_0\beta \hat{w}(q)=0$. This leads to a very simple linear density dependence of the onset temperature: $k_BT=|\hat{w}(k=q)|\rho_0$, where the marginally stable wavenumber $q$ at onset is determined by the condition
\begin{equation}
\frac{d\hat{w}(k)}{dk}\bigg|_{k=q}=0.
\end{equation}
For the 2D GEM-8 system the onset wavenumber $q\approx5.26/R$. Since $\hat{w}(k=q)\approx -0.294 \epsilon R^2$ is independent of the density, the linear instability threshold is a straight line in the phase diagram:
\begin{equation}\label{eq:lin_instab_line}
\frac{k_BT}{\epsilon}\approx 0.294\rho_0R^2
\end{equation}
{(Fig.~\ref{fig:phase_diag}). In addition, the binodals along which the liquid and crystal phases coexist in thermodynamic equilibrium also appear to be straight lines in the phase diagram. This is not obvious {\it a priori} because the binodal calculation requires that one first obtains the crystal density profile, which is a highly nonlinear problem. However, fitting the numerically obtained binodals with a straight line proves to be an excellent approximation (Fig.~\ref{fig:phase_diag}). For example, for the GEM-8 fluid we find that} the binodal for the crystal state at coexistence is given by $k_BT/\epsilon\approx 0.314\rho_0R^2$ and that of the liquid is $k_BT/\epsilon\approx 0.339\rho_0R^2$. Thus, when the temperature $k_BT/\epsilon=1$, the density of the liquid at coexistence is $\rho_0R^2\approx1/0.339=2.95$ while that of the crystal is $\rho_0R^2\approx1/0.314=3.18$.

{At thermodynamic coexistence, at the temperature $T=T_{\rm coex}$ (and the chemical potential $\mu=\mu_{\rm coex}$), a front between the crystal and the liquid state is stationary. However, on decreasing the temperature below $T=T_{\rm coex}$ (or increasing $\mu$ above $\mu=\mu_{\rm coex}$)}, the liquid state is no longer the equilibrium state. For a shallow quench, the liquid state remains linearly stable but a crystal can still grow if it is nucleated: only a crystal seed that is larger than the critical size grows and the interface (front) between the two phases advances at a well-defined speed $v_{\rm nl}$ determined by nonlinear processes \cite{AWTK14, HN00}. This (pushed) front propagation was studied in detail in Ref.\ \cite{AWTK14} for the GEM-4 model. At $T=T_{\rm coex}$, the front speed $v_{\rm nl}=0$; as the temperature $T$ decreases below $T_{\rm coex}$ the speed $v_{\rm nl}$ increases with increasing quench depth $|T-T_{\rm coex}|$.

If the quench is to a temperature $T<T_{\rm sp}$, where $T_{\rm sp}$ is the temperature determined by Eq.\ (\ref{eq:lin_instab_line}) at which the uniform liquid becomes linearly unstable (i.e.\ the spinodal), then front propagation via linear processes is possible, with the speed $v$ determined by the marginal stability analysis described in \S\ref{sec:4}. However, as can be seen from Eqs.\ (\ref{eq:v_smalld}), (\ref{eq:k_i_smalld}) and (\ref{eq:k_r_approx}), $v=0$ at $T=T_{\rm sp}$ since $\omega_m=0$ at $T_{\rm sp}$. As the temperature is decreased below $T_{\rm sp}$, $v$ increases but remains less than $v_{\rm nl}$ for small $|T-T_{\rm sp}|$. In this regime the front remains a pushed front even though the liquid is already unstable \cite{AWTK14,HN00,CC97}. However, the speed $v$ increases faster than $v_{\rm nl}$ with decreasing $T$ resulting in a crossover in speeds at temperature $T=T_{\rm x}$, where $T_{\rm x}<T_{\rm sp}<T_{\rm coex}$. At $T_{\rm x}$, the two speeds are equal, $v=v_{\rm nl}$, but for a sufficiently deep quench $v>v_{\rm nl}$, and for these temperatures ($T<T_{\rm x}$) it is the linear process that determines how the crystal state advances into the unstable liquid \cite{AWTK14,HN00,CC97}.

\begin{figure}[b]
\centering
\includegraphics[width=0.49\columnwidth]{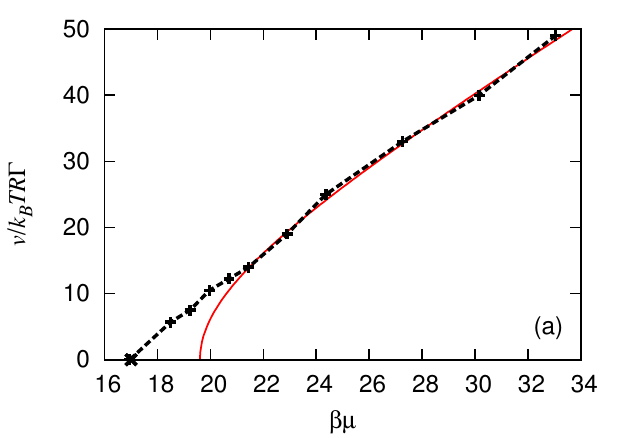}
\includegraphics[width=0.49\columnwidth]{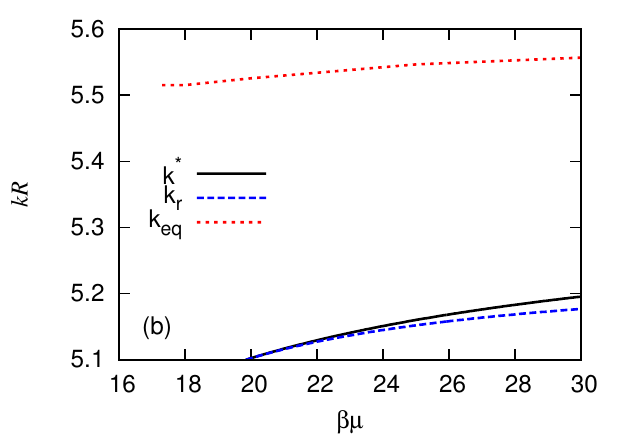}
\caption{(a) The front speed $v$ as a function of the chemical potential $\mu$ for the GEM-4 fluid with temperature $k_BT/\epsilon=1$. The solid line is the result of the marginal stability calculation while the symbols connected with a dashed line summarize the results from numerical simulations of the 2D DDFT equations. (b) The wavenumber $k_r$ of the density modulation selected by the moving front, the wavenumber $k^*$ deposited behind the front (both calculated from the marginal stability condition), and the wavenumber $k_{\rm eq}$ of the equilibrium crystal. The difference between $k^*$ and $k_{\rm eq}$ implies that rearrangements behind the front are inevitable as the system seeks to minimise its free energy.}
   \label{fig:front_speed}
\end{figure}

{The variation of the speed of the crystallisation front with increasing chemical potential $\mu$ (at fixed temperature) is analogous to that described above for decreasing temperature (at fixed chemical potential). The metastable uniform liquid becomes linearly unstable at $\mu_{\rm sp}>\mu_{\rm coex}$ and for $\mu>\mu_{\rm sp}$ front propagation via linear processes is possible. However, it is only when $\mu>\mu_{\rm x}>\mu_{\rm sp}$ that linear processes govern} {the propagation of the front and the front speed is determined by the marginal stability result. In Fig.\ \ref{fig:front_speed}(a) (see also Fig.~4 of Ref.\ \cite{AWTK14}) we show for a GEM-4 fluid with temperature $k_BT/\epsilon=1$ that for $\beta\mu>\beta\mu_{\rm x}\approx21$ the front speed obtained from solving the DDFT equations numerically in 2D does indeed agree precisely with the speed $v$ predicted by the marginal stability analysis. Figure \ref{fig:front_speed}(b) (see also Fig.~6 of Ref.\ \cite{AWTK14}) compares the wavenumber $k^*$ deposited behind a front generated by a deep quench ($\mu>\mu_{\rm x}$, equivalently $T<T_{\rm x}$) with the wavenumber $k_{\rm eq}$ corresponding to the equilibrium crystal lattice spacing. Our aim below is to explore the consequences of the dramatic difference between $k^*$ and $k_{\rm eq}$ revealed in the figure for the subsequent evolution of the solid phase, and to demonstrate that it is responsible for the inevitable presence of defects and disorder.}

\section{Structures formed after a quench}
\label{sec:7}

In Fig.\ \ref{fig:log_rho_T1_0} we display a series of snapshots of the density profile calculated for a 2D GEM-8 fluid with bulk density $\rho_0R^2=5$, quenched to the temperature $T^*=k_BT/\epsilon=1$, where the uniform fluid is linearly unstable [cf.\ Eq.\ (\ref{eq:lin_instab_line})]. The times are given in Brownian time units, $t^*\equiv k_BT\Gamma R^2t$, where $t^*\sim1$ is the time it takes for a particle to diffuse a distance $\sim R$. Solidification is initiated by imposing at time $t=0$ small amplitude random fluctuations along the line $x=0$ on an otherwise uniform density profile with periodic boundary conditions. This produces a pair of solidification fronts propagating out to the left and to the right from this line. These fronts can be observed in the top left time $t^*=1$ density profile in Fig.\ \ref{fig:log_rho_T1_0}. Ahead of the front one observes stripe-like oscillations in the liquid, as predicted by the marginal stability analysis (see Refs.\ \cite{AWTK14, ARTK12} for other examples of similar fronts). Behind this stripe-like precursor the crystal grows but contains defects, a consequence of the mismatch between the dynamically generated length scale $2\pi/k^*$ and the equilibrium length scale $\ell$: to lower its free energy and form the equilibrium crystal structure, the system must perform significant rearrangements. These rearrangements occur locally, generating defects behind the propagating crystallisation front. By time $t^*=2$, the two fronts have collided owing to the periodic boundary conditions (see top right panel in Fig.\ \ref{fig:log_rho_T1_0}) and as time proceeds the crystal density distribution continues to rearrange, removing defects in order to lower the free energy. By time $t^*=100$ (bottom right) the system is largely occupied by a well-ordered hexagonal crystal with the  correct equilibrium length scale $\ell$. However, even after such a long time, there is still a region where the crystal orientation is not aligned with the crystal in the centre of the system -- i.e.\ two different crystal grains remain present in the system, with defects remaining on the grain-boundary between these two. By `grain boundaries' we refer to regions where regular hexagonal ordering is absent as occurs along the boundaries of neighbouring regions of misaligned hexagonal ordering (the `grains'). 
\begin{figure}
\centering
\includegraphics[width=0.49\columnwidth]{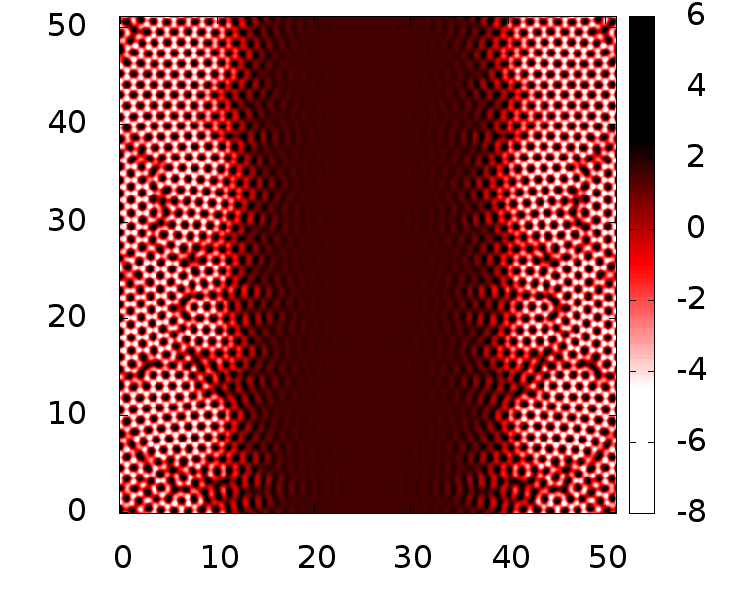}
   \includegraphics[width=0.49\columnwidth]{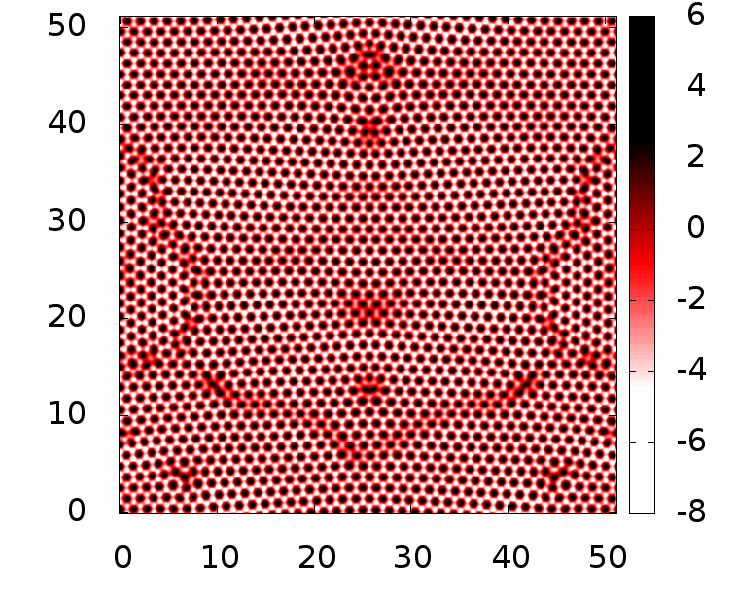}
   
   \includegraphics[width=0.49\columnwidth]{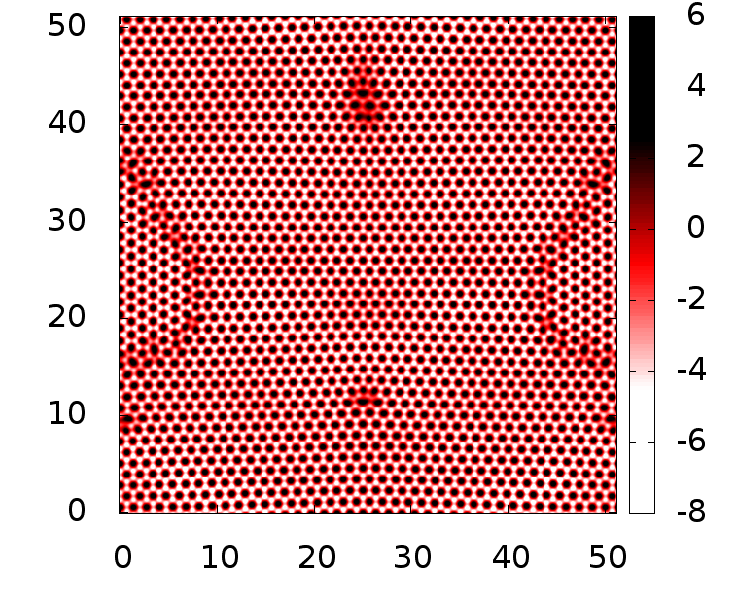}
   \includegraphics[width=0.49\columnwidth]{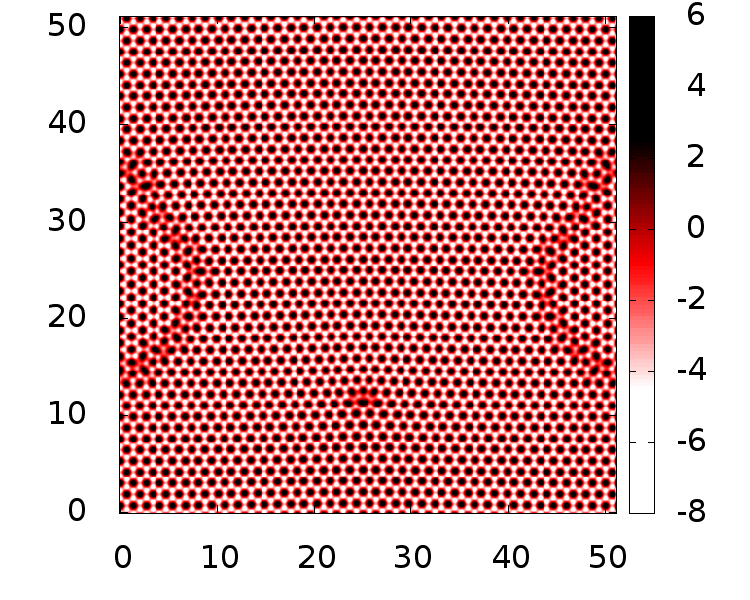}
   \caption{The logarithm of the 2D density profile, $\ln(\rho({\bf r},t)R^2)$, at times $t^*=1$, 2, 10 and 100 (going from top left to bottom right) obtained from DDFT for $\rho_0 R^2=5$ and temperature $k_BT/\epsilon=1$. Plotting the logarithm of the density, rather than the density itself, allows one to see the grain boundaries and defects more clearly. The solidification front was initiated at time $t^*=0$ along the line $x=0$, or equivalently $x=51.2R$.}
   \label{fig:log_rho_T1_0}
\end{figure}

\begin{figure}
\centering
\includegraphics[width=0.49\columnwidth]{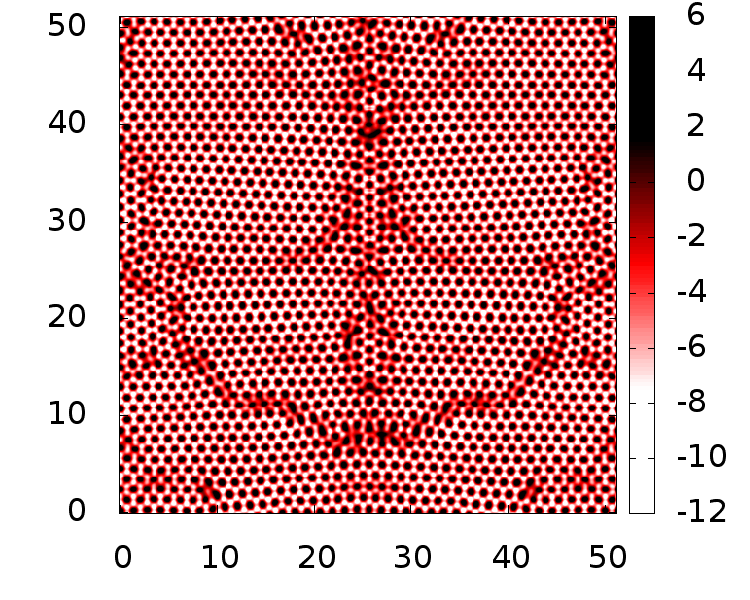}
   \includegraphics[width=0.49\columnwidth]{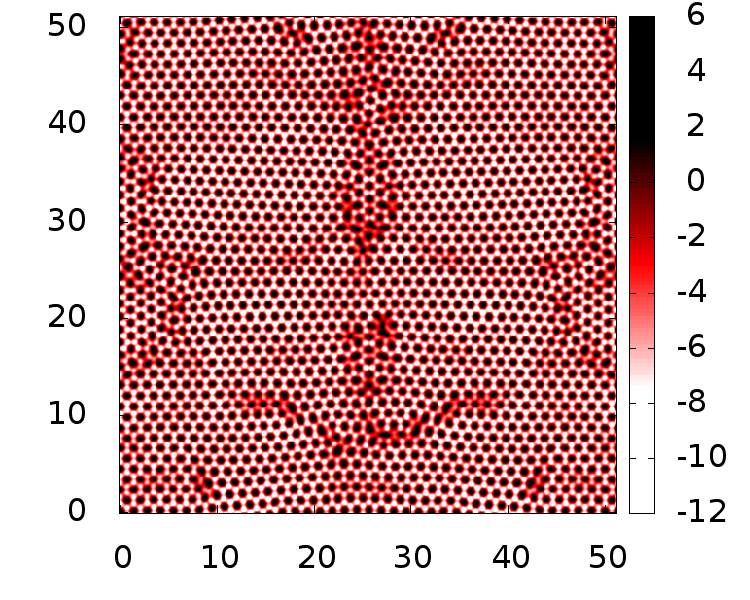}
   
   \includegraphics[width=0.49\columnwidth]{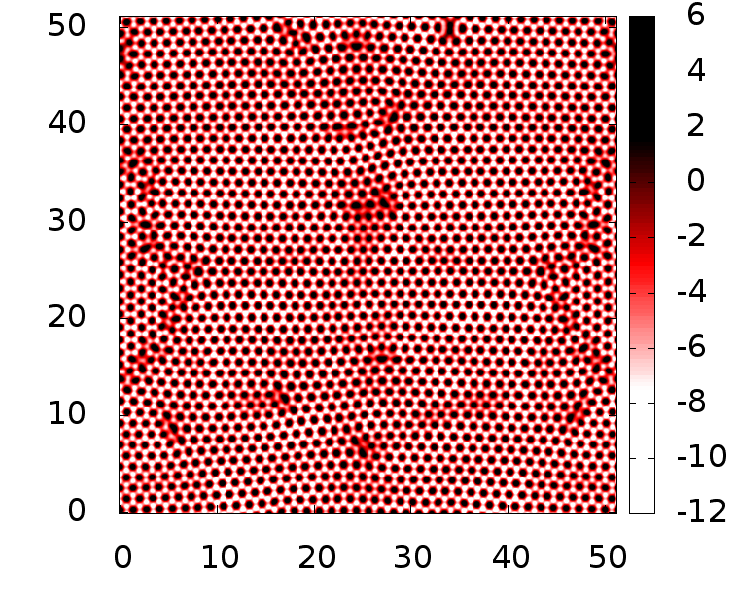}
   \includegraphics[width=0.49\columnwidth]{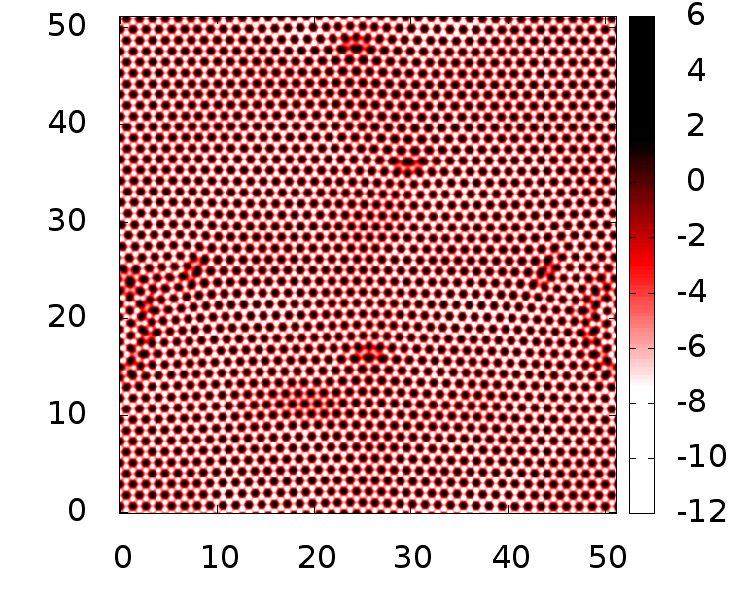}
   \caption{Same as Fig.\ \ref{fig:log_rho_T1_0}, but for the lower temperature $k_BT/\epsilon=0.75$.}
   \label{fig:log_rho_T0_75}
\end{figure}

In Fig.\ \ref{fig:log_rho_T0_75} we display a series of density profile snapshots for a system with the same density as in Fig.\ \ref{fig:log_rho_T1_0}, but here quenched to a lower temperature, $T^*=k_BT/\epsilon=0.75$. In this case the solidification front speed $v$ is faster, so the two fronts have already propagated across the system by the time $t^*=1$. In keeping with the prediction in Eq.\ (\ref{eq:k_star}), the mismatch between $2\pi/k^*$ and the equilibrium length scale $\ell$ increases with quench depth (i.e.\ with the parameter $\omega_m$) and so more defects and disorder are produced in the system. Over time these mostly anneal out, although some persist for very long times.

\begin{figure}
\centering
\includegraphics[width=0.49\columnwidth]{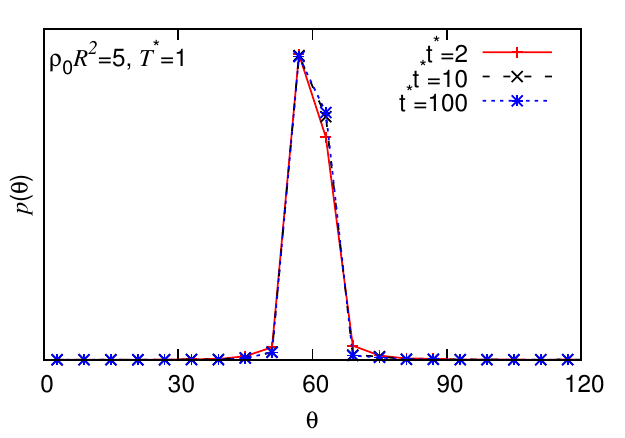}
\includegraphics[width=0.49\columnwidth]{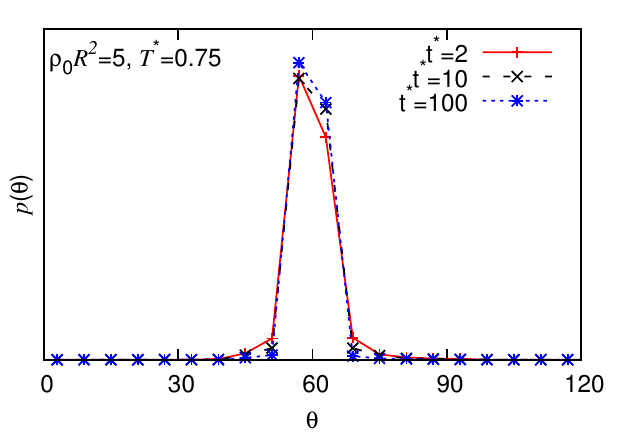}
   \caption{The bond angle distribution $p(\theta)$, calculated from Delauney triangulation on the density profiles displayed in Figs.\ \ref{fig:log_rho_T1_0} and \ref{fig:log_rho_T0_75}. The distribution exhibits a single peak at $60^\circ$ due to the predominance of hexagonal ordering although shoulders on either side of this peak are present at early times due to the disorder in the system. These decrease over time, as the system rearranges to remove defects and lower the free energy.}
   \label{fig:bond_angle_1}
\end{figure}

To quantify the degree of ordering in the density profiles, we perform a Delauney triangulation \cite{delauney, RATK12, ARTK12, AWTK14} on the points corresponding to all the maxima in the density profile. We then calculate the distribution function $p(\theta)$ of the bond angles $\theta$ in these triangles. Perfect hexagonal ordering results in a Delauney triangulation consisting entirely of equilateral triangles, corresponding to a single sharp peak at $60^\circ$ in $p(\theta)$. In Fig.\ \ref{fig:bond_angle_1} we display $p(\theta)$ at times $t^*=2$, 10 and 100 after the quench, corresponding to the results displayed in Figs.\ \ref{fig:log_rho_T1_0} and \ref{fig:log_rho_T0_75}. We do indeed see a single peak in $p(\theta)$, centred at $60^\circ$. However, the peak also exhibits `shoulders' on either side, particularly for short times after the quench, indicating the presence of strain and disorder at early times. These shoulders diminish over time, but never completely disappear. Comparing the results for the temperature $T^*=1$ (left plot in Fig.\ \ref{fig:bond_angle_1}) with those for $T^*=0.75$ (right plot in Fig.\ \ref{fig:bond_angle_1}), we observe that for shallower quenches ($T^*=1$) the distribution $p(\theta)$ at time $t^*=2$ is closer to the later time equilibrium distribution than for deeper quenches ($T^*=0.75$). These results confirm and quantify the impression derived from the 2D density profiles, i.e.\ that the deeper the quench, the greater the disorder created in the system and the more rearrangements that are required to equilibrate the system. This is also what one would expect on the basis of Eq.\ (\ref{eq:k_star}).

\section{A binary mixture}\label{sec:8}

The results in the previous section (see also Refs.\ \cite{ARTK12, AWTK14}) show that systems composed of only one species of particle, when quenched, are generally able to rearrange to form a well-ordered crystalline solid, even if initially there is disorder in the system. However, since binary mixtures are far more likely to be glass formers, we now consider a binary mixture of soft GEM-8 particles, interacting via the pair potential [cf.\ Eq.\ (\ref{eq:pair_pot})]:
\begin{equation}
w_{ij}(r)=\epsilon_{ij}e^{-(r/R_{ij})^8} \,,
\end{equation}
where the indices $i,j=1,2$ label the two different species of particles. We choose the pair potential parameters $\epsilon_{12}/\epsilon_{11}=1$, $\epsilon_{22}/\epsilon_{11}=1.5$, $R_{12}/R_{11}=1$ and $R_{22}/R_{11}=1.3$, corresponding to the case where the potentials $w_{11}(r)$ and $w_{12}(r)$ are identical, but the potential $w_{22}(r)$ between pairs of species 2 particles is stronger and of longer range. We use DDFT to determine the dynamics of the system following a quench of the uniform liquid to temperatures where it is linearly unstable and a hexagonal crystal is the equilibrium state. We approximate the free energy of the binary mixture using a two-component generalisation of Eqs.\ (\ref{eq:grand_pot}) -- (\ref{eq:F_ex}), input into the DDFT for mixtures -- i.e.\ the generalisation of Eqs.\ (\ref{eq:continuity}) and (\ref{eq:DDFT_current}) -- as described in detail in Ref.\ \cite{Archer05}. For simplicity, we set the mobility coefficients of the two species to be equal, i.e.\ $\Gamma_1=\Gamma_2=\Gamma$.
\begin{figure}
   \centering   
   \includegraphics[width=0.49\columnwidth]{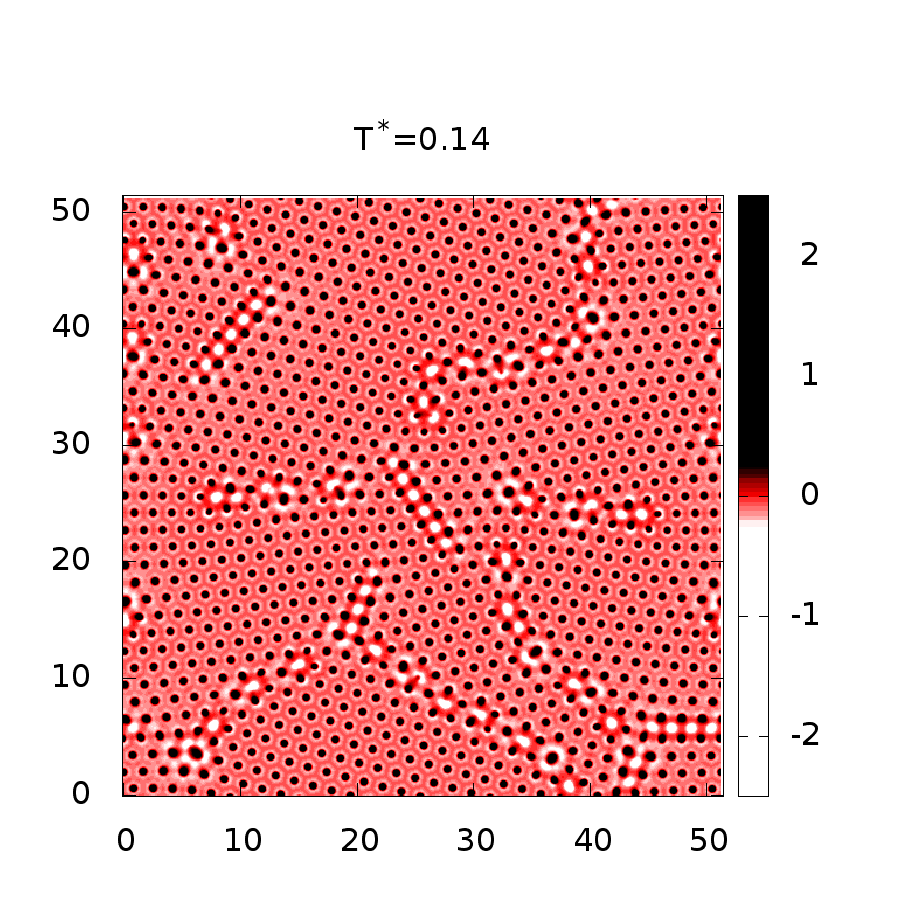}
   \includegraphics[width=0.49\columnwidth]{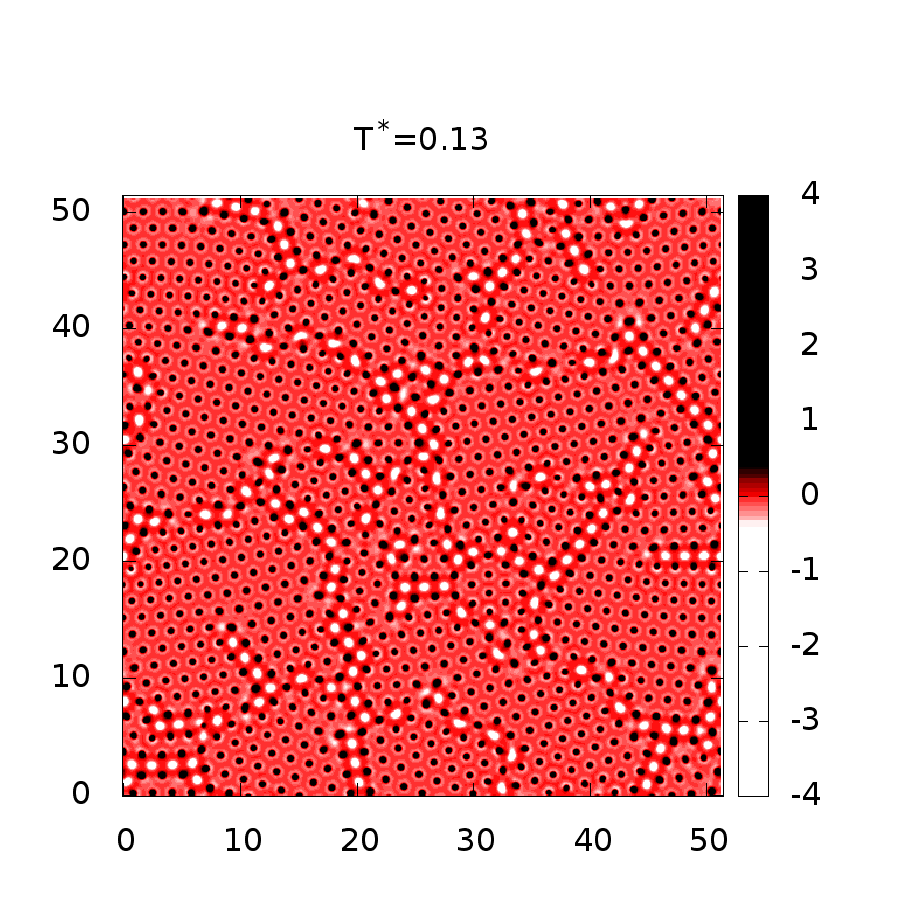}
   
   \includegraphics[width=0.49\columnwidth]{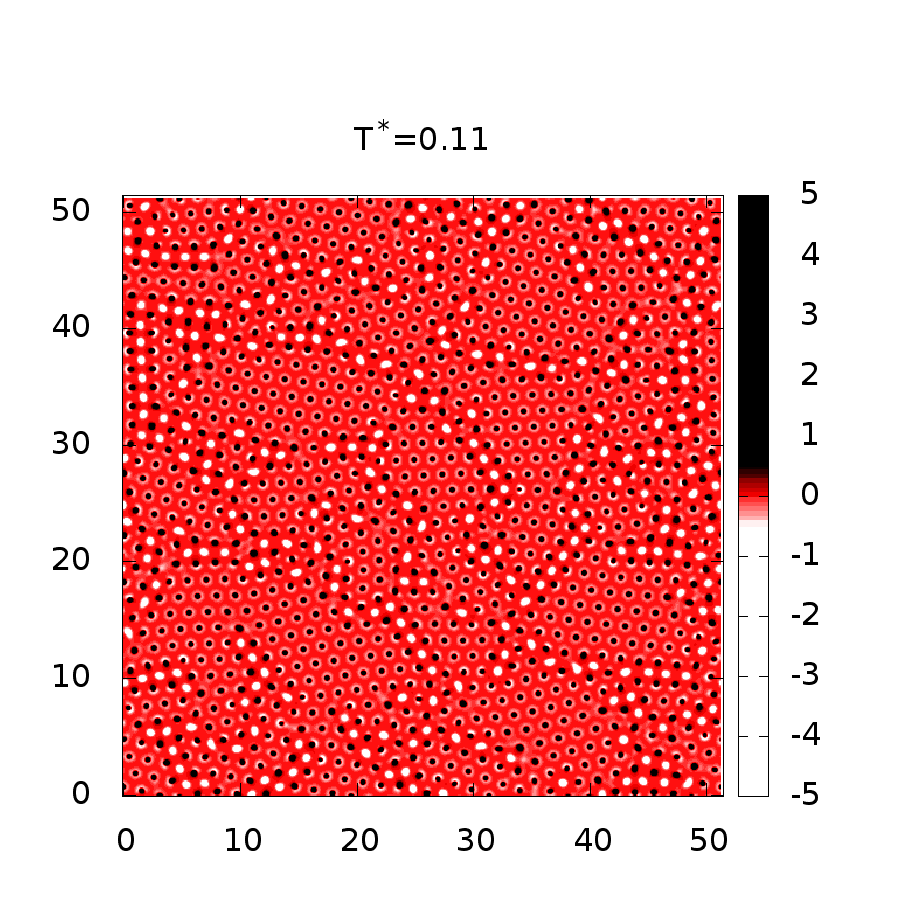}
   \includegraphics[width=0.49\columnwidth]{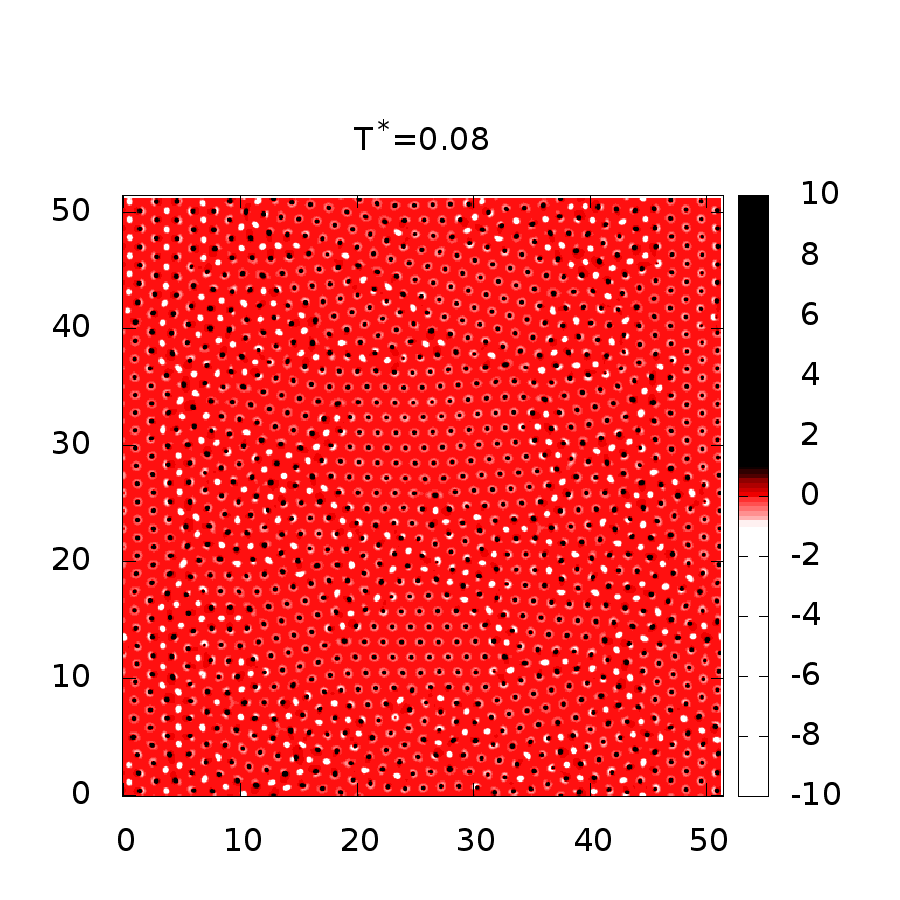}
   \caption{The density difference $[\rho_1({\bf r})-\rho_2({\bf r})]R_{11}^2$ after time $t^*=100$ (when the profiles are almost stationary) obtained from DDFT with average densities $\bar{\rho}_1R_{11}^2=\bar{\rho}_2R_{11}^2=0.2$ and the pair potential parameters $\epsilon_{12}/\epsilon_{11}=1$, $\epsilon_{22}/\epsilon_{11}=1.5$, $R_{12}/R_{11}=1$ and $R_{22}/R_{11}=1.3$. The dimensionless temperatures $T^*=k_BT/\epsilon_{11}$ are given above each figure. Plotting this quantity reveals the location of the species 2 particles (regions coloured white) which aggregate at the defects and grain boundaries. The grain size increases with the temperature $T^*$.}
   \label{fig:mixture}
\end{figure}

\begin{figure}[t]
\sidecaption[t]
\includegraphics[width=0.6\columnwidth]{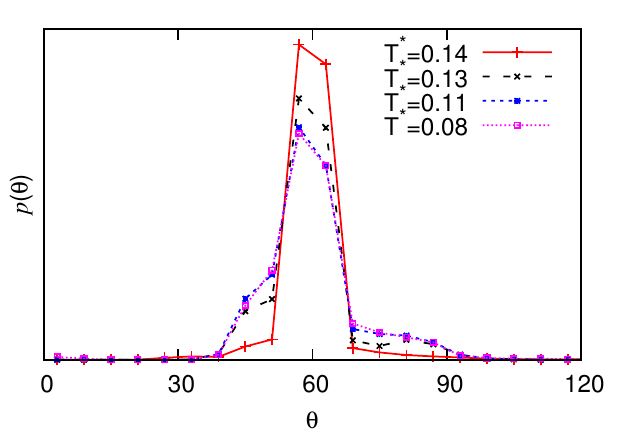}
\caption{The distribution $p(\theta)$ of bond angles obtained from Delauney triangulation, corresponding to the density profiles in Fig.\ \ref{fig:mixture}. Note that as the temperature decreases $p(\theta)$ becomes broader, indicating that there is greater disorder in the system.}
\label{fig:bond_op}
\end{figure}

In Fig.\ \ref{fig:mixture} we display results for the final equilibrium state obtained from quenching a 50:50 binary fluid with average densities $\bar{\rho}_1R_{11}^2=\bar{\rho}_2R_{11}^2=0.2$ down to various different temperatures, $T^*=k_BT/\epsilon_{11}$. The linear instability threshold $T_{\rm sp}$ for these densities is at $T^*=0.149$. For temperatures below this value, the uniform liquid is linearly unstable. Above this value, the uniform liquid is linearly stable and for the crystal to form, it must be nucleated. Note that at all the temperatures for which we display results in Fig.\ \ref{fig:mixture}, the density profiles of the two species cease evolving after a time of $t^*\approx 100-200$. In Fig.\ \ref{fig:mixture} we plot the quantity $[\rho_1({\bf r})-\rho_2({\bf r})]R_{11}^2$ and regions where $\rho_1>\rho_2$ are coloured black while regions where $\rho_1<\rho_2$ are coloured white. Regions coloured red are where either (i) both densities are small, or (ii) the two densities are equal in magnitude, but not necessarily small. It is worth observing that around most of the black density peaks there is a pale `ring'. This is because wherever there is a peak in the density of species 1, there is also normally a peak in the density of species 2. The peaks in $\rho_2({\bf r})$ are, however, broader than the peaks in $\rho_1({\bf r})$ but lower in height, hence the `ring'. They are lower in height because many of the species 2 particles gather at the grain boundaries; these are the white peaks in the profiles in Fig.\ \ref{fig:mixture}. The eye can easily pick out the grain boundaries in these plots, from the lines of white peaks. We also observe that the density peaks at higher temperatures are broader than those at the lower temperatures (cf.\ Lindemann's criterion for melting \cite{HM}). Comparing the different profiles in Fig.\ \ref{fig:mixture} we see that the deeper the quench, the smaller the size of the crystalline grains and the more disorder is present, since there are more grain boundaries in the system. This observation is supported by the results in Fig.\ \ref{fig:bond_op} which displays the bond angle distribution $p(\theta)$ obtained from Delauney triangulation. For quenches to lower temperatures, $p(\theta)$ is broader, indicating that there is greater disorder in the system.

These results show that deep quenches to temperatures below the linear instability threshold result in the formation of states that are highly disordered -- a consequence of the fact that the propagating solidification front deposits behind it a dynamically generated length scale that differs from that of the equilibrium hexagonal crystal, thereby introducing defects into the system. In contrast to the one-component system discussed in \S\ref{sec:7}, binary mixtures of particles are not able to rearrange as easily as in a monodisperse system and so the disorder created by the solidification front persists for a very long time (indeed indefinitely, within the present simple mean-field DFT treatment).

\section{System with competing crystal structures}\label{sec:9}

We now present results for a different GEM-8 binary mixture, which has pair potential parameters $\epsilon_{11}=\epsilon_{12}=\epsilon_{22}$, $R_{12}/R_{11}=1$ and $R_{22}/R_{11}=1.5$. This particular system was first considered in Ref.\ \cite{AWTK14}, and we provide here additional detail. For fixed (sufficiently high) total density $\bar{\rho}\equiv\bar{\rho}_1+\bar{\rho}_2$, where $\bar{\rho}_1$ and $\bar{\rho}_2$ are the average densities of species 1 and 2, respectively, the system exhibits several different equilibrium crystal structures depending on the concentration $\Phi\equiv\bar{\rho}_1/\bar{\rho}$ of species 1. Examples of these are displayed in Fig.\ \ref{fig:binary_mixture_profiles}. All the profiles displayed correspond to local minima of the free energy, but we have not checked whether they are the global minima at the given state points. At low values of $\Phi$ we observe a hexagonal crystal, such as that displayed in Fig.\ \ref{fig:binary_mixture_profiles}(a). However, for larger values of $\Phi$, the system can form a binary square crystal structure, such as that displayed in Fig.\ \ref{fig:binary_mixture_profiles}(c). Alternatively, it can form a binary hexagonal lattice structure, such as that displayed in Figs.\ \ref{fig:binary_mixture_profiles}(b) and \ref{fig:binary_mixture_profiles}(d). For high values of $\Phi$ the system forms a different simple hexagonal lattice, an example of which is displayed in Fig.\ \ref{fig:binary_mixture_profiles}(e). In this hexagonal crystal the minority species particles occupy the same lattice sites as the majority species particles, in contrast to the lattice structures in (b)--(d). See Ref.\ \cite{AWTK14} for further details.

\begin{figure}[t]
%\centering
\includegraphics[width=0.3\columnwidth]{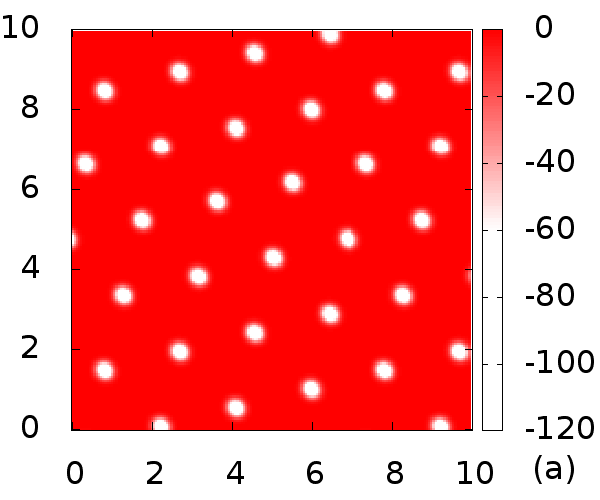}
\includegraphics[width=0.3\columnwidth]{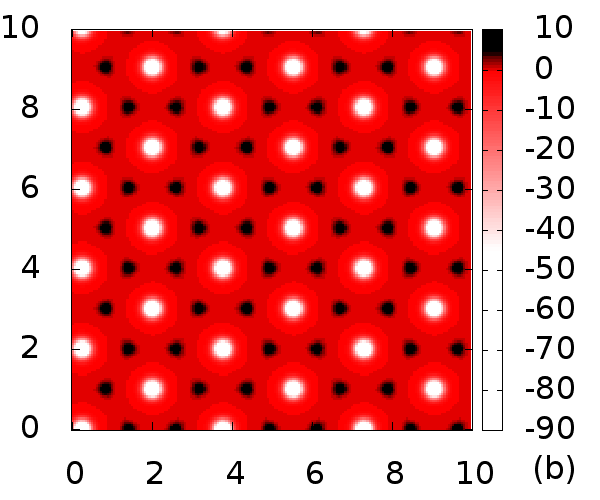}
\includegraphics[width=0.3\columnwidth]{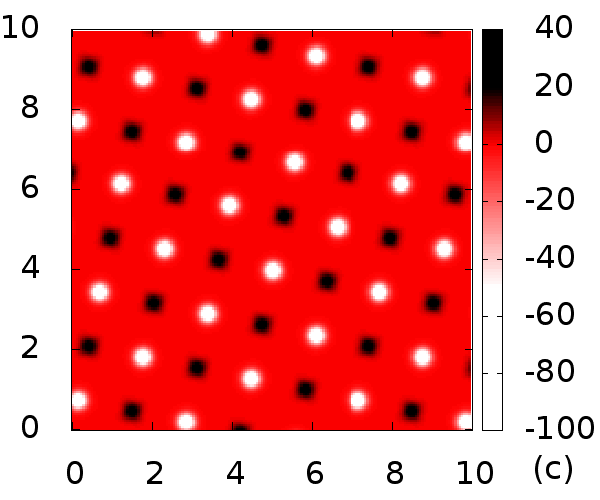}

\includegraphics[width=0.3\columnwidth]{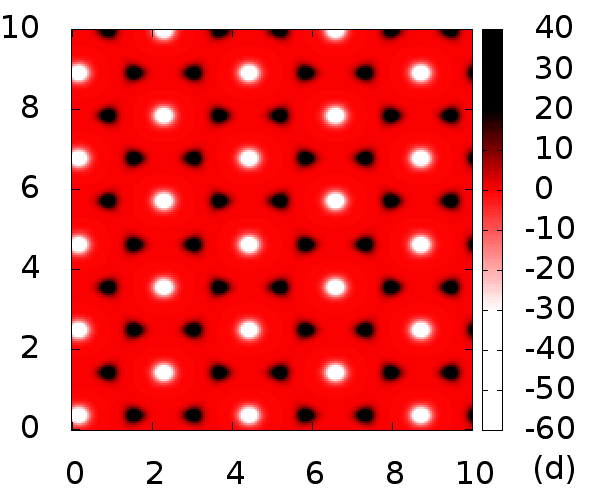}
\includegraphics[width=0.3\columnwidth]{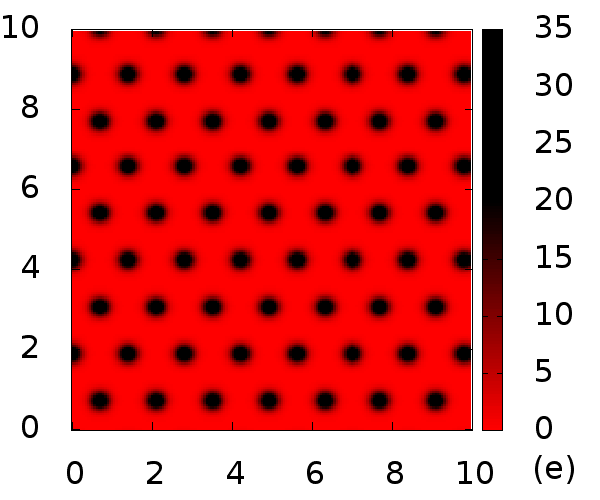}
\includegraphics[width=0.39\columnwidth]{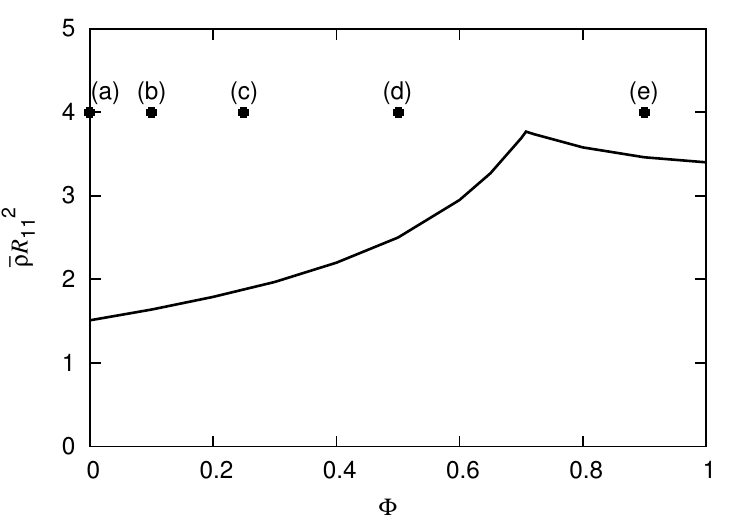}

\caption{Equilibrium crystal structures for the GEM-8 mixture with $\beta \epsilon_{ij}=\beta \epsilon=1$ for all $i,j=1,2$, $R_{22}/R_{11}=1.5$, $R_{12}/R_{11}=1$ for concentrations (a) $\Phi=0$, (b) 0.1, (c) 0.25, (d) 0.5, (e) 0.9 and average total density $\bar{\rho} R_{11}^2=4$. We plot the quantity $[\rho_1({\bf r})-\rho_2({\bf r})]R_{11}^2$, so regions where $\rho_1({\bf r})>\rho_2({\bf r})$ are coloured black while those where $\rho_1({\bf r})<\rho_2({\bf r})$ are coloured white. All profiles correspond to local minima of the free energy, but we have not checked whether they correspond to global minima at the given state points. In (f) we display the linear stability threshold density as a function of concentration $\Phi$, for fixed temperature $k_BT/\epsilon=1$.}
   \label{fig:binary_mixture_profiles}
\end{figure}

\begin{figure}
\centering
\includegraphics[scale=.65]{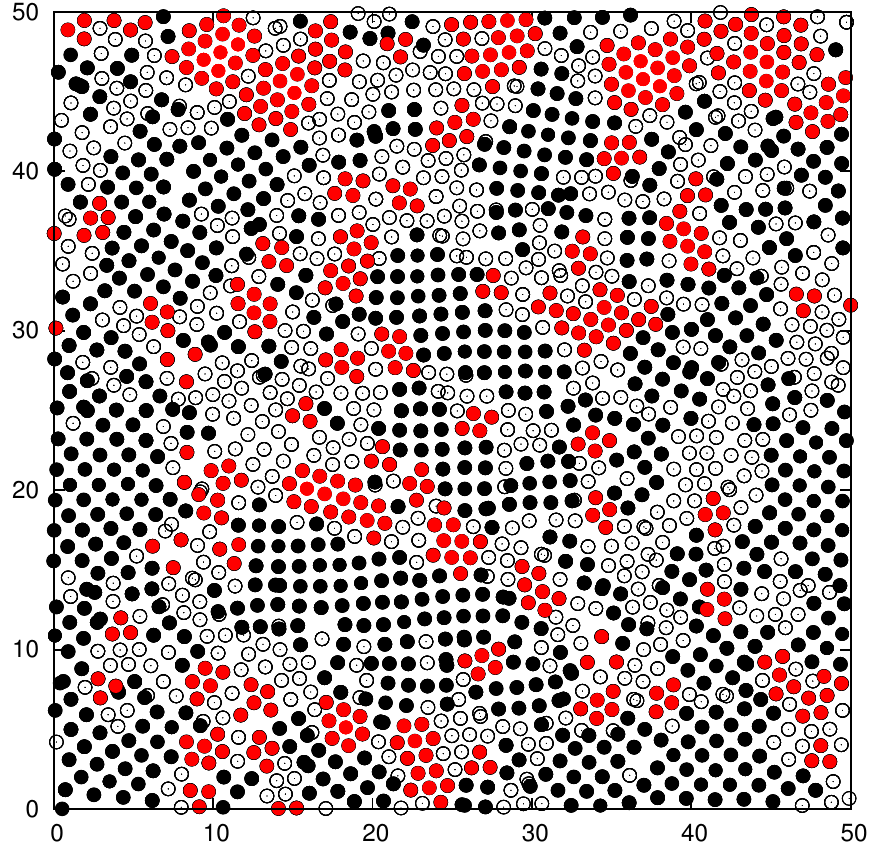}
\includegraphics[scale=.65]{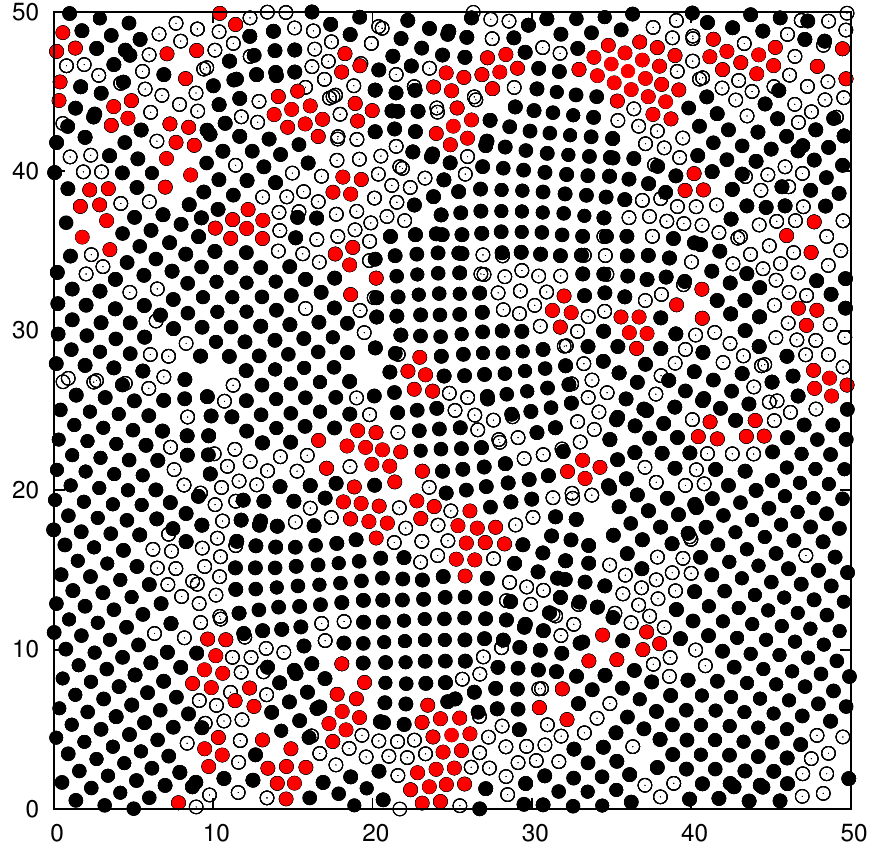}

\includegraphics[scale=.65]{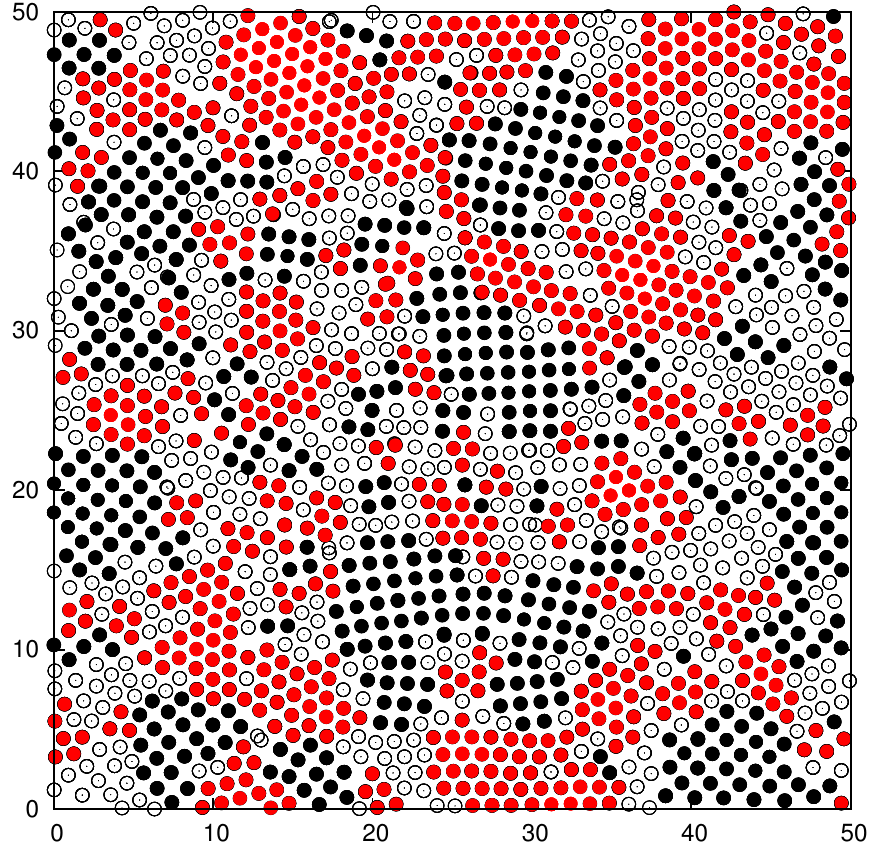}
\includegraphics[scale=.65]{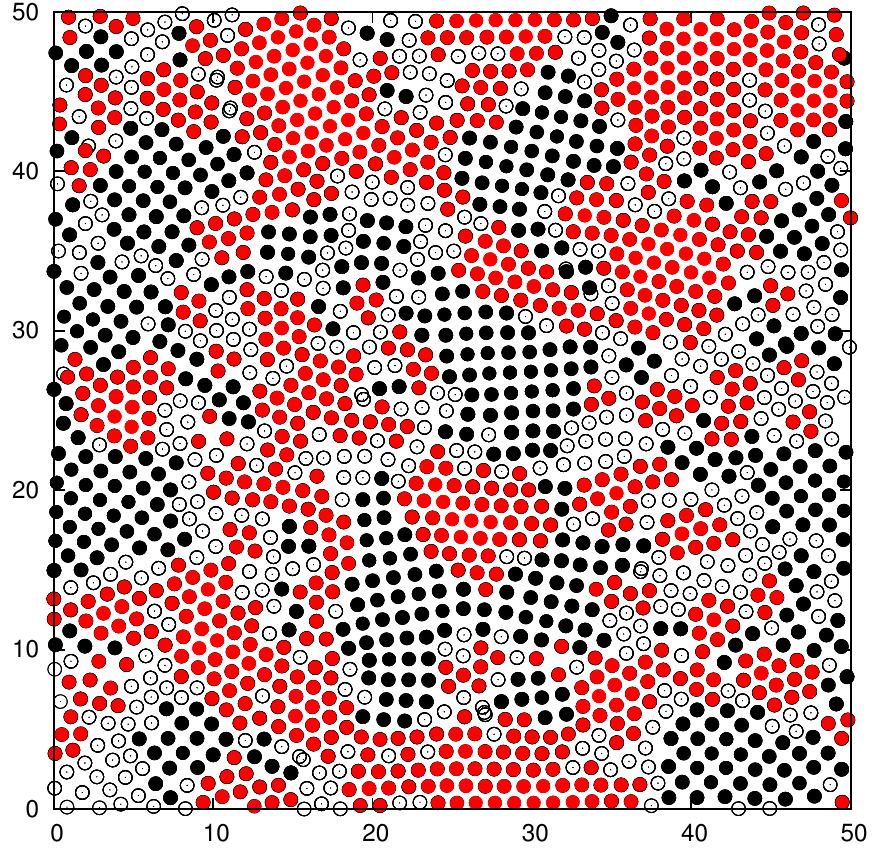}

\includegraphics[scale=.65]{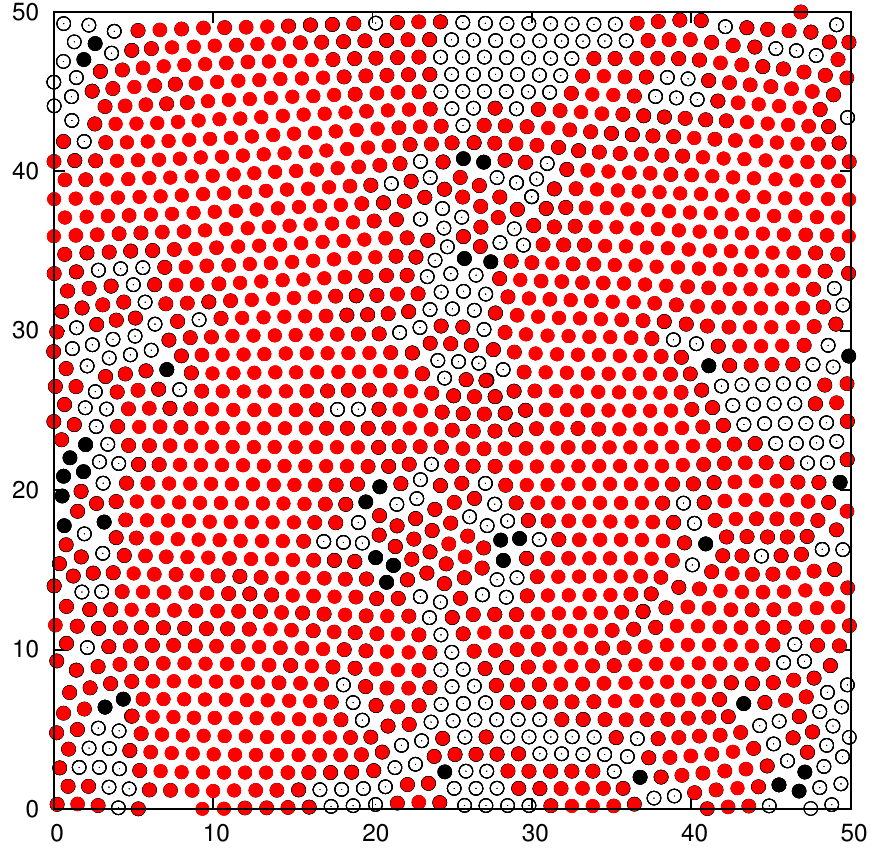}
\includegraphics[scale=.65]{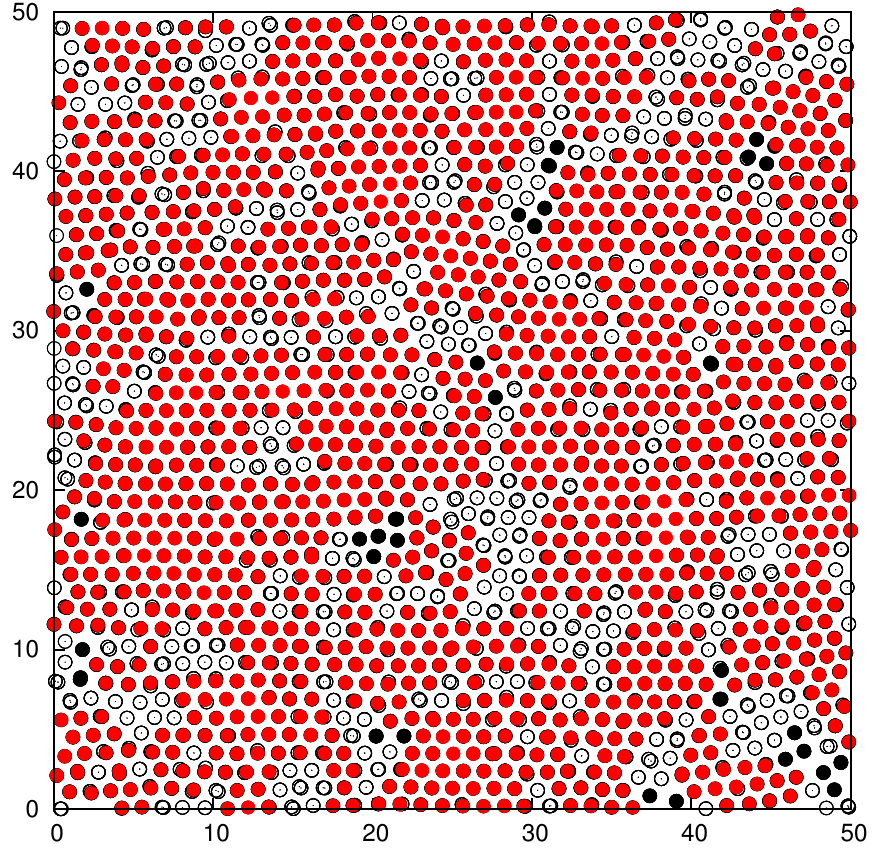}
\caption{The peaks in the density profile formed by a solidification front initiated along the line $x = 25$ at time $t = 0$. The system is a GEM-8 mixture with $\beta \epsilon_{ij}=1$ for all $i,j = 1,2$, $R_{22}/R_{11} = 1.5$ and $R_{12}/R_{11} = 1$ and average total density $(\bar{\rho}_1+\bar{\rho}_2)R_{11}^2\equiv\bar{\rho} R_{11}^2=8$. The results in the top row are for $\Phi=0.25$, in the middle row for $\Phi=0.5$ and the bottom row for $\Phi=0.75$. In each case the plot on the left is for an early time $t^*=1.6$, shortly after the solidification front has exited the domain and before the structure has had time to relax, while that on the right is for a later time (top right and bottom right are for $t^*=40$, whilst the middle right is for $t^*=400$). The density maxima are colour-coded according what kind of triangle they belong to in a Delauney triangulation: right-angled are black, equilateral are red and scalene are open circles. Portions of the hexagonal crystal is red, whilst the competing square crystal structure is black.}
\label{fig:spots}
\end{figure}

We use DDFT to determine the structures that are formed when the uniform liquid mixture is quenched. A solidification front is initiated along the line $x/R_{11}=25$ at time $t=0$ by adding a small random value along this line. The fact that there are several competing crystal structures, in conjunction with the fact that we quench to state points that are far from the linear instability threshold (i.e.\ we perform a deep quench) means that the structures that are formed are highly disordered. 

In Fig.\ \ref{fig:spots} we present results from quenches for three systems with total density $\bar{\rho}R_{11}^2=8$ but different concentrations: $\Phi=0.25$ (top), $\Phi=0.5$ (middle) and $\Phi=0.75$ (bottom).\footnote{The early time profile displayed in the top right panel in Fig.~12 of Ref.~\cite{AWTK14} shows the corresponding plot for $\Phi=0.25$ instead of $\Phi=0.5$ as labelled there. This error is corrected here in Fig.~\ref{fig:spots}.} Instead of displaying the density profiles calculated using DDFT, we plot the locations of all maxima in the total density profile $\rho({\bf r})\equiv\rho_1({\bf r})+\rho_2({\bf r})$ exceeding $50R_{11}^{-2}$ at the peak. In addition, we colour-code these points, according to the local structure. The criteria for this are obtained via a Delauney triangulation on the points. Points where the local ordering is square are coloured black, points where there is hexagonal order are coloured red and the density peaks with neither local ordering are shown as open circles.\footnote{The specific criteria for deciding to which subset a given density peak belongs is as follows: After performing the Delauney triangulation on the set of peaks, we consider each triangle. The corner angles are $\theta_1$, $\theta_2$ and $\theta_3$. The triangle is defined as equilateral if $|\theta_i-\theta_j|<5^{\circ}$ for all pairs $i,j=1,2,3$. The vertices of these triangles are coloured black. Triangles are defined as right-angled if for the largest angle $\theta_1$ we have $|\theta_1-90^\circ |<5^{\circ}$ and for the other two angles $|\theta_2-\theta_3|<5^{\circ}$. The vertices of these triangles are coloured red. All remaining vertices which fall into neither of these categories are displayed as open circles.} In Fig.\ \ref{fig:spots} we display on the left the structure formed at time $t^*=1.6$, i.e., shortly after the solidification front has exited the box, and also the structure at a much later time (at time $t^*=40$ for $\Phi=0.25$ and $\Phi=0.75$ and the time $t^*=400$ for the case $\Phi=0.5$ displayed on the  middle right in Fig.\ \ref{fig:spots}). The structures that are formed are all highly disordered.

\begin{figure}%[t]
\centering
\includegraphics[scale=1.]{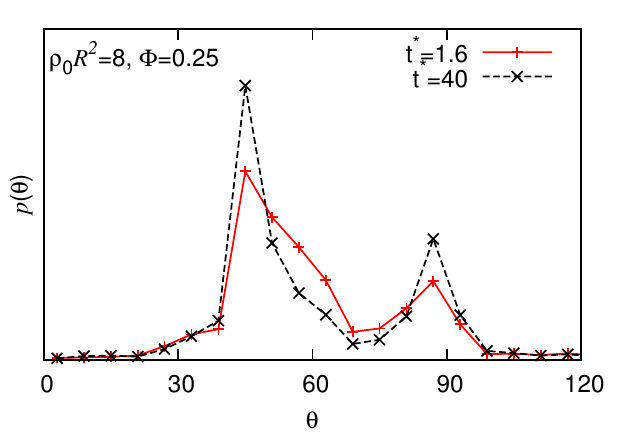}

\includegraphics[scale=1.]{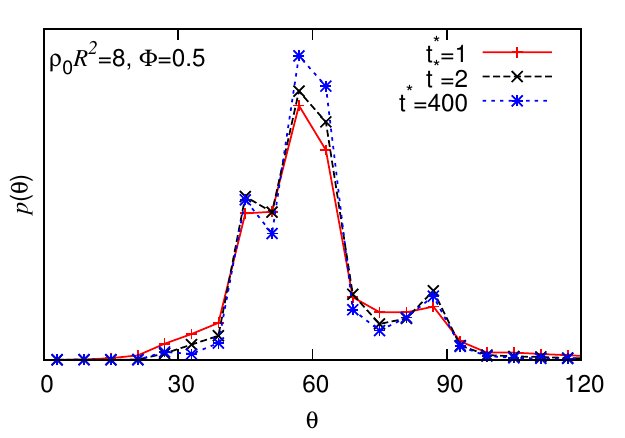}

\includegraphics[scale=1.]{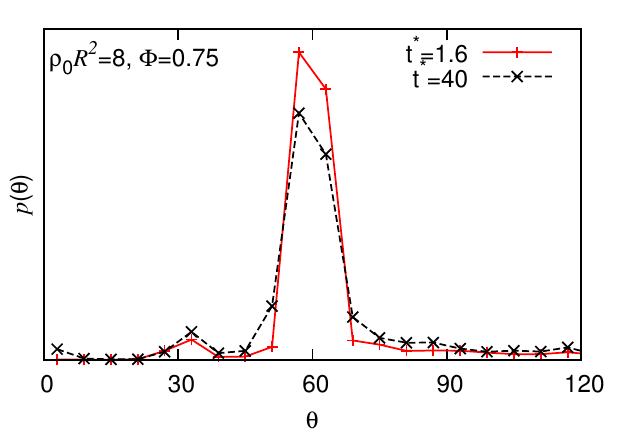}
\caption{The distribution $p(\theta)$ of bond angles obtained from Delauney triangulation, corresponding to the results in Fig.\ \ref{fig:spots}, which are for $\bar{\rho} R_{11}^2=8$ and concentration $\Phi=0.25$ (top), $\Phi=0.5$ (middle) and $\Phi=0.75$ (bottom). In the top plot, the peaks in the distribution at 45$^\circ$ and 90$^\circ$ show that the system is largely composed of square ordered local structure. In contrast, the bottom plot, for $\Phi=0.75$, has just one main peak in the distribution at 60$^\circ$ showing that the system is largely composed of hexagonal ordered local structure. In the middle plot, which is for $\Phi=0.5$, $p(\theta)$ has peaks in the distribution at 45$^\circ$, 60$^\circ$ and 90$^\circ$ showing that the system contains both square ordered local structure and also hexagonal local ordering, as one can also see from Fig.\ \ref{fig:spots}.}
\label{fig:bond_op_2}
\end{figure}

In the $\Phi=0.25$ case displayed along the top row in Fig.\ \ref{fig:spots} the structure that forms initially is a mixture of square and hexagonal ordering, but with a predominance of squares. There are also regions containing neither structure, largely on the grain boundaries. Over time, the regions of square ordering grow at the expense of the regions with hexagonal ordering. The density profiles continue to evolve very slightly after $t^*=40$, but the displayed structure is very similar in its statistical properties to the final equilibrium structure.

The bond angle distribution $p(\theta)$ corresponding to the $\Phi=0.25$ profiles is displayed in the top panel of Fig.\ \ref{fig:bond_op_2}. This distribution has two main peaks at 45$^\circ$ and 90$^\circ$, reflecting significant square ordering in the system. Such squares yield right-angle triangles in the Delauney triangulation. Initially, at time $t^*=1.6$, these peaks are rather broad, but over time they become sharper, reflecting the presence of growing domains of a well-ordered square crystal. The hexagonal ordering in the system is reflected by the fact that the peak in $p(\theta)$ at 45$^\circ$ has a `shoulder' on it, extending out to 60$^\circ$. This shoulder diminishes in height over time, but does not completely disappear, reflecting the fact that small regions of hexagonal ordering persist.

In the middle row of Fig.\ \ref{fig:spots}, we display the density peaks formed after quenching a uniform liquid with concentration $\Phi=0.5$. Once again the structure that is formed contains regions of both square and hexagonal ordering. The size of both of these types of domains grows over time. In the middle right of Fig.\ \ref{fig:spots} we display the density peaks at time $t^*=400$, after which the density profiles cease to evolve. The corresponding bond angle distribution $p(\theta)$ is displayed in the middle panel of Fig.\ \ref{fig:bond_op_2}. This distribution has three peaks at 45$^\circ$, 60$^\circ$ and 90$^\circ$, reflecting the presence of a mixture of squares and hexagons. These peaks are rather broad, reflecting the significant disorder in the system.

In the bottom row of Fig.\ \ref{fig:spots}, we display the structure formed from quenching a concentration $\Phi=0.75$ uniform liquid. We see that hexagonal ordering dominates, a fact confirmed by the large main peak in $p(\theta)$ located at 60$^\circ$ displayed in the lower panel of Fig.\ \ref{fig:bond_op_2}. What is particularly remarkable about these $\Phi=0.75$ results is that the peak at 60$^\circ$ is actually sharper at the early times ($t^*=1.6$) than later times ($t^*=40$): over time the peak broadens! This is due to the fact that in this case the solidification front produces modulations with wavenumber $k^*$ that is close to the wavenumber for the hexagonal crystal structure. However, these do not match exactly so that the hexagonal crystal that is initially formed is strained. Over time, the system lowers the free energy by {\em introducing} defects which alleviate the strain. These defects lead in turn to the broadening of the peak in $p(\theta)$ at 60$^\circ$.

\section{Concluding remarks}\label{sec:10}

In this paper we have discussed a mechanism that results in the formation of disordered structures, when a liquid is deeply quenched to temperatures where the thermodynamic equilibrium state is a well-ordered crystal. This occurs because solidification fronts in deeply quenched liquids propagate via a mechanism that generates periodic density modulations in the system with wavelength that is not necessarily the same as the wavelength required for an equilibrium crystal. The wavelength mismatch means that the formation of a well-ordered equilibrium crystal state requires significant rearrangements after the front has passed. In monodisperse one-component systems, these rearrangements should generally be possible; this is certainly the case in the model fluid studied here. However, for polydisperse systems or multi-component mixtures, such as the binary mixtures studied here, these rearrangements are frustrated and in some cases hindered by the fact that there is a variety of particle sizes in the system.

We should emphasise that the front propagation mechanism focused on in this paper operates only when the quench is sufficiently deep, to temperatures below the crossover temperature $T_{\rm x}$. Only for $T<T_{\rm x}$ do solidification fronts propagate via the linear mechanism, with the speed $v$ and wavenumber $k^*$ determined by linear considerations. Above $T_{\rm x}$ the front speed is determined by nonlinear considerations and in this regime the structure formed behind the front is generally much better ordered.

The results presented here are for rather simple 2D model systems composed of soft particles. Nevertheless, we believe that the mechanism that we describe should be rather general, although much further work is required to determine the nature of crystallisation fronts in other deeply quenched systems, and in particular to determine whether one can reach the regime where the fronts propagate via the linear mechanism that we describe. In other systems, it may be the case that the speed $v$ never overtakes $v_{\rm nl}$. For example, it is not known whether this regime is physically accessible for particles with a hard core; it may be that this regime only arises at densities near or even beyond random close-packing.

As a final point, we should mention that the mechanism described here is not the only means of introducing disorder as liquids solidify. Other mechanism include: (i) Defects created by impurities. (ii) Different grains, with defects on the grain boundaries, generated in the nucleation regime when growing crystals with different orientation nucleated at different points in the system collide. (iii) Defects introduced by crystal growth under the influence of external forces or shear. (iv) Disordered materials produced in shallow quenches where the growing crystal forms dendritic type structures (diffusion limited growth), leading to the formation of crystal grains and defects.

\begin{acknowledgement}
A.J.A. and U.T. thank the Center of Nonlinear Science (CeNoS) of the University of M\"{u}nster for recent support of their collaboration. M.C.W. is supported by an EPSRC studentship. The work of E.K. was supported in part by the National Science Foundation under Grant No. DMS-1211953.
\end{acknowledgement}

\end{document}